\documentclass[epjST]{svjour} 

\usepackage{mathtools}
\usepackage{amsmath} 

\usepackage{hyperref}

\begin{document}

\title{Progress in the numerical studies of the type IIB matrix model}

\author{%
Konstantinos N. Anagnostopoulos\inst{1}     \fnmsep\thanks{\email{konstant@mail.ntua.gr}}
Takehiro        Azuma\inst{2}               \fnmsep\thanks{\email{azuma@mpg.setsunan.ac.jp}}
Kohta           Hatakeyama\inst{3}          \fnmsep\thanks{\email{khat@post.kek.jp}}
Mitsuaki        Hirasawa\inst{4}            \fnmsep\thanks{\email{Mitsuaki.Hirasawa@mib.infn.it}}
Yuta            Ito\inst{5}                 \fnmsep\thanks{\email{y-itou@tokuyama.ac.jp}}
Jun             Nishimura\inst{3,6}         \fnmsep\thanks{\email{jnishi@post.kek.jp}}
Stratos         Kovalkov Papadoudis\inst{1} \fnmsep\thanks{\email{sp10018@central.ntua.gr}}
Asato           Tsuchiya\inst{7}            \fnmsep\thanks{\email{tsuchiya.asato@shizuoka.ac.jp}}                                                                                            }
\institute{%
Physics Department, School of Applied Mathematical and Physical Sciences, National Technical University of Athens, Zografou Campus, Zografou, GR-15780, Greece                           \and
Institute for Fundamental Sciences, Setsunan University, 17-8 Ikeda Nakamachi, 17-8 Ikeda Nakamachi, 572-8508, Osaka, Japan                                                              \and
Theory Center, Institute of Particle and Nuclear Studies, High Energy Accelerator Research Organization (KEK), 1-1 Oho, Tsukuba, 305-0801, Ibaraki, Japan                                \and
Sezione di Milano-Bicocca, Istituto Nazionale di Fisica Nucleare (INFN), Piazza della Scienza 3, Milano, I-20126, Italy                                                                  \and
National Institute of Technology, Tokuyama College, Gakuendai, Shunan, 745-8585, Yamaguchi, Japan                                                                                        \and
Department of Particle and Nuclear Physics, School of High Energy Accelerator Science, Graduate University for Advanced Studies (SOKENDAI), 1-1 Oho, Tsukuba, 305-0801, Ibaraki, Japan   \and
Department of Physics, Shizuoka University, 836 Ohya, Suruga-ku, 422-8529, Shizuoka, Japan                                                                                                   }
\abstract{The type IIB matrix model, also known as the IKKT model, has been proposed as a promising candidate for a non-perturbative formulation of superstring theory. Based on this proposal, various attempts have been made to explain how our four-dimensional space-time can emerge dynamically from superstring theory. In this article, we review the progress in numerical studies on the type IIB matrix model. We particularly focus on the most recent results for the Euclidean and Lorentzian versions, which are obtained using the complex Langevin method to overcome the sign problem. We also review the earlier results obtained using conventional Monte Carlo methods and clarify the relationship among different calculations.
}


\maketitle

\section{Introduction}\label{intro}

The type IIB matrix model, also known as the IKKT model \cite{9612115}, is regarded as one of the most promising candidates for a
non-perturbative formulation of superstring theory. The model is defined by dimensionally reducing ten-dimensional ${\cal N}=1$
super Yang--Mills theory to zero dimensions. Therefore, space-time does not exist a priori in this model. We interpret the
eigenvalues of the bosonic matrices as the space-time coordinates, and hence, space-time is generated dynamically from the degrees
of freedom of the matrices~\cite{Aoki:1998vn}. Since superstring theory is defined in ten dimensions, it is important to understand
how our four-dimensional space-time emerges by studying this model.

Various attempts have been made to address this question. In the Lorentzian version of the type IIB matrix model, the indices are
contracted by the metric $\eta_{\mu \nu} = \textrm{diag} (-1,1,\dots,1)$, and the action has the SO(9,1) symmetry. The bosonic
action is unbounded from below, and this is why no one has dared to study the Lorentzian version numerically for a long
time. Instead, the efforts have been focused on the Euclidean version
\cite{9811220,0003208,0005147,0108041,1009_4504,1108_1534,1306_6135,1509_05079,1712_07562,2002_07410}, which is defined by making a
Wick rotation with respect to the temporal direction, and contracting the indices by the Euclidean metric $\delta_{\mu \nu} =
\textrm{diag} (1,1,\dots,1)$.

The Euclidean version has the SO(10) rotational symmetry instead of the SO(9,1), and it is amenable to numerical simulations
because the partition function is finite without any cutoffs \cite{9803117,0103159}. However, it suffers from a severe sign
problem, which appears after integrating out the fermions. The complex Pfaffian (determinant in the simplified four or
six-dimensional SUSY models) plays a central role in the spontaneous symmetry breaking (SSB) of the SO(10) rotational
symmetry~\cite{0003223,Nishimura:2000wf}. In models where there are no fermionic degrees of freedom, like the bosonic model, or the
Pfaffian is real positive, like in the four-dimensional SUSY model, there is no SSB of the rotational
symmetry~\cite{9811220,0003208,Ambjorn:2001xs}. There is no SSB, either, in the phase-quenched model, which omits the complex phase
of the Pfaffian~\cite{1509_05079}. Thus, to study the SSB of the rotational symmetry, one should take into account the complex
phase, for instance by reweighting, in which case the important configurations are generally different between the original model
and its phase-quenched version, leading to a severe overlap problem. To reduce this problem, the factorization method
\cite{0108041,1009_4504,1108_1534,1306_6135,1509_05079} simulates a constrained system, in which the expectation value of the phase
factor is calculated to determine the true vacuum. The results are consistent with the SSB pattern of SO(10) to SO(3) predicted
using the Gaussian expansion method (GEM) \cite{1007_0883,1108_1293}. While this is an interesting dynamic property, its relevance to our
four-dimensional space-time is unclear.

This observation led to the Monte Carlo simulation of the Lorentzian version of the type IIB matrix model
\cite{1108_1540,1312_5415,1506_04795,1904_05914}. The problem of the unbounded bosonic action was solved using separate cutoffs in
the temporal and spatial directions~\cite{1108_1540}.  Although the Pfaffian is real, the model has a severe sign problem due to
the bosonic part of the action $S_{\textrm{b}}$, which appears with a factor $e^{i S_{\textrm{b}}}$ in the partition function. To
avoid it, the authors in Ref.~\cite{1108_1540} used an approximation, and they found that {\it three} out of nine spatial
dimensions start to expand after a critical time. This moment, which results from the dynamics of the model, may be identified as
the birth of the universe. Later works~\cite{1312_5415,1506_04795} computed the expansion rate of the universe numerically, which
starts with an exponential-law expansion at early times, followed by a power-law expansion at late times. However, the authors in
Ref.~\cite{1904_05914} found that, because of the approximation, the effect is due to the domination of almost three--dimensional
configurations with a singular Pauli-matrix structure.

The complex Langevin method (CLM), a stochastic process for complexified variables, is a promising method for simulating a system
that suffers from the sign problem \cite{Parisi:1983mgm,Klauder:1983sp}. Recently, the CLM has attracted a lot of attention because
the condition for the equivalence to the original path integral has been clarified
\cite{0912_3360,1101_3270,1504_08359,1606_07627}. The authors of Refs.~\cite{1712_07562,2002_07410} applied the CLM to the
Euclidean version of the type IIB matrix model, and they reproduced the SSB of the SO(10) symmetry to SO(3). The application of the
CLM to the Lorentzian version \cite{1904_05919,2201_13200,2205_04726} may elucidate the space-time structure that emerges when we
exclude the approximation to avoid the sign problem. While this is still an ongoing work, there are some preliminary results
\cite{2205_04726} that look quite promising.

In this article, we review the exciting progress in numerical studies on the type IIB matrix model. The rest of this article is
organized as follows. In Sec.~\ref{typeIIBmodel}, we provide a review of the type IIB matrix model as a promising nonperturbative
formulation of superstring theory. In Sec.~\ref{old_IKKT_numerical}, we review the numerical studies on the Euclidean version using
the factorization method and those on the Lorentzian version using the approximation to avoid the sign problem. In
Sec.~\ref{CLM_sec} we provide an overview of the CLM and discuss its application to the Euclidean and Lorentzian versions.
Sec.~\ref{sec13} is devoted to a summary and an outlook. See
Refs.~\cite{Klinkhamer:2020xoi,Brahma:2021tkh,Steinacker:2021yxt,Brahma:2022dsd,Battista:2022hqn,Karczmarek:2022ejn} for other
recent studies on the type IIB matrix model, which suggest possible applications to cosmology.  The reader may also consult earlier
reviews~\cite{Aoki:1998bq,Nishimura:2003rj,Azuma:2004vs,Steinacker:2010rh,Nishimura:2012xs,Nishimura:2014vza,Ydri:2017ncg,Nishimura:2020blu}
and the references therein.

%
\section{The type IIB matrix model}\label{typeIIBmodel}
In this section, we review the type IIB matrix model, focusing on how the space-time appears in this model. We also discuss the reason why it can be regarded as a nonperturbative formulation of superstring theory.

\subsection{Symmetries of the type IIB matrix model}
The action of the type IIB matrix model \cite{9612115} is given by
\begin{eqnarray}
 S &=& S_{\textrm{b}} + S_{\textrm{f}}   \ ,      
 \label{IKKT_action}
 \\
 S_{\textrm{b}} &=& - \frac{1}{4g^2} \textrm{tr} ([A_{\mu}, A_{\nu}] [A^{\mu}, A^{\nu}] )\ , \label{IKKT_boson} \\
 S_{\textrm{f}} &=& - \frac{1}{2g^2} \textrm{tr} \left( {\bar\psi}_\alpha  (\Gamma^{\mu})_{\alpha\beta} [A_{\mu}, \psi_\beta] \right) \ , \label{IKKT_fermion}
\end{eqnarray}
where the $A_{\mu}$ ($\mu=0,1,2,\dots,9)$ are bosonic $N\times N$ Hermitian matrices, the $\psi_\alpha$ ($\alpha=1,\dots, 16$) are fermionic $N\times N$ Hermitian matrices, and the $\Gamma^{\mu}$ are the ten-dimensional gamma matrices.  The action is obtained by dimensionally reducing the ten-dimensional $\mathcal{N}=1$ super Yang--Mills theory (SYM) to zero dimensions. Hence, space-time does not exist a priori in this model.  Note, however, that the action has ten-dimensional Lorentz symmetry; namely, it is invariant under the SO(9,1) transformation in which $A_{\mu}$ and $\psi_{\alpha}$ are transformed as a vector and a Majorana-Weyl spinor, respectively.  Since $g$ is merely a scale factor that can be absorbed by rescaling $A_\mu$ and $\psi_\alpha$, in what follows, we set $g^2 N = 1$ without loss of generality.

One can easily see that the action is also invariant under the following transformations:
\begin{align}
&  \left\{
  \begin{array}{lcl}
  \delta^{(1)}A_{\mu} &=&i\bar{\epsilon}_1\Gamma_{\mu}\psi \ ,
    \\
  \delta^{(1)}\psi &=&\frac{i}{2} \Gamma^{\mu\nu}[A_{\mu},A_{\nu}]\epsilon_1 \ ,
  \end{array}
  \right.
  \label{SUSY1}\\
&  \left\{
  \begin{array}{lcl}
    \delta^{(2)}A_{\mu}&=& 0 \ ,
    \\
    \delta^{(2)}\psi &=&\epsilon_2 1_N \ ,
  \end{array}
  \right.
    \label{SUSY2} \\
&  \left\{
  \begin{array}{lcl}
    \delta_{\textrm{transl}} A_{\mu} &=&c_{\mu}1_N \ ,
    \\
    \delta_{\textrm{transl}} \psi &=& 0 \ ,
  \end{array}
  \right.
  \label{translation} \\
&    \left\{
  \begin{array}{lcl}
    \delta_{\textrm{gauge}} A_{\mu} &=& i[\lambda, A_{\mu}] \ ,
    \\
    \delta_{\textrm{gauge}} \psi &=& i[\lambda,\psi] \ ,
  \end{array}
  \right.
    \label{gauge}
\end{align}
where $\epsilon_1$ and $\epsilon_2$ are Grassmann odd parameters with Majorana-Weyl spinor indices, $c_{\mu}$ is a vector parameter, and $\lambda$ is a parameter that is a $N \times N$ Hermitian matrix.  The transformations (\ref{SUSY1}) and (\ref{gauge}) are nothing but the dimensional reduction of the supersymmetry transformation and the gauge transformation in ten-dimensional $\mathcal{N}=1$ SYM, respectively.

We denote the generators of the transformations (\ref{SUSY1}), (\ref{SUSY2}) and (\ref{translation}) by $Q^{(1)}$, $Q^{(2)}$ and $P_{\mu}$, respectively, and define $\tilde{Q}^{(1)}$ and $\tilde{Q}^{(2)}$ by
\begin{align}
\tilde{Q}^{(1)} & =   Q^{(1)} + Q^{(2)}  \ , \nonumber\\
\tilde{Q}^{(2)} & = i(Q^{(1)} - Q^{(2)}) \ .
\end{align}
Then, using the equation of motion for $\psi$
\begin{align}
  \Gamma^{\mu}[A_{\mu},\psi]=0 \ ,
  \label{eq:psi-EOM}
\end{align}
we find the relation
\begin{align}
[\bar{\epsilon}_1\tilde{Q}^{(i)}, \bar{\epsilon}_2\tilde{Q}^{(j)}] = -2 \delta^{ij}\bar{\epsilon}_1 \Gamma^{\mu}\epsilon_2 P_{\mu}\, ,
\label{N=2 SUSY}
\end{align}
up to terms proportional to the gauge transformation (\ref{gauge}).

Note that, if Eq.~(\ref{translation}) is identified with the translation, then Eq.~(\ref{N=2 SUSY}) is the algebra of ten-dimensional $\mathcal{N}=2$ supersymmetry\footnote{This is an on-shell supersymmetry because the equation of motion \eqref{eq:psi-EOM} is used.}. This suggests that the eigenvalues of $A_{\mu}$ can be interpreted as ten-dimensional coordinates~\cite{Aoki:1998vn}. Thus, space-time is generated dynamically from the degrees of freedom of the bosonic matrices. Since $\mathcal{N}=2$ supersymmetry in ten dimensions is maximal, any theory with this symmetry must include gravitons, provided that the theory is unitary and has a massless spectrum. Therefore, the fact that the type IIB matrix model possesses the symmetries generated by  $Q^{(1)}$, $Q^{(2)}$ and $P_{\mu}$ suggests strongly that it includes gravity.

\subsection{Connection to the type IIB superstring theory}
In this subsection, we discuss the connection of the type IIB matrix model to type IIB superstring theory, which is a perturbative formulation of superstring theory. 

First, the matrix model can be viewed as a matrix regularization of the worldsheet formulation
of type IIB superstring theory \cite{9612115}. If one considers the Green--Schwarz action of the IIB superstring in the Schild gauge~\cite{Schild:1976vq}
\begin{equation}
\label{k01}
S_{\mbox{\scriptsize Schild}} = \int\, d^2\sigma \sqrt{g} \left(\frac{1}{4}\{X_\mu,X_\nu\}^2 - \frac{i}{2}\bar\psi\Gamma^\mu\{X_\mu,\psi\}\right)\, ,
\end{equation}
then we can use the semiclassical correspondence, valid in the large--$N$ limit,
\begin{equation}
\label{k02}
\left\{\,\, , \,\,\right\} \to -i\left[ \,\, , \,\, \right]\, \qquad \int\,d^2\sigma\,\sqrt{g} \to {\rm tr} 
\end{equation}
to obtain the action (\ref{IKKT_action}). 

Second, while the worldsheet formulation is the first quantization of superstrings,  the matrix model is expected to give a second quantization because multiple string worldsheets appear naturally 
as block-diagonal configurations, where each block represents a single string worldsheet. The fluctuations of off-diagonal blocks are considered to represent the interaction between the worldsheets. 

Third, the long-distance behavior of the interaction between D-branes in
type IIB superstring theory is reproduced by the one-loop calculation in the matrix model \cite{9612115}. 

Finally, by identifying the matrix model ``Wilson loops'' 
\begin{equation}
\label{k03}
W_C = {\rm tr} \prod_{n=1}^M \exp\left\{i\epsilon(k_n^\mu A_\mu+\bar\lambda_n\psi) \right\}\, ,
\end{equation}
with the regularized creation operators of strings, the light-cone string field theory for type IIB superstrings is derived from the 
Schwinger-Dyson equations for the loops under a few reasonable assumptions \cite{9705128,Aoki:1998bq}. In the above formula, $k_n^\mu$ is identified with the momentum density on the (fundamental) string, and $\lambda_n$ with the fermionic sources. The short-distance cutoff $\epsilon$ is expected to appear dynamically, and vanish in the large--$N$ limit, in a way that the $W_C$ remain finite. This double scaling limit depends nontrivially on the nonperturbative dynamics of the model \cite{0003208,Anagnostopoulos:2001cb}. 

We conclude that the matrix model can reproduce the perturbative expansion in type IIB superstring theory to all orders. In this manner, the matrix model has a direct connection to a perturbative formulation of 
superstring theory. In this connection, we see again that the eigenvalues of the matrices are identified with the space-time coordinates.

Let us recall the so-called string duality, which states that each perturbative formulation of superstring theory corresponds to a point in the moduli space of the unique superstring theory. Assuming that this is true, one should be able to obtain the whole theory by starting
from any point in the moduli space using a theory defined in a non-perturbative manner. Thus, the direct connection to type IIB superstring theory, together with the fact that the model is defined without relying on perturbation theory, ensures that the type IIB matrix model can be
a non-perturbative formulation of superstring theory.

\section{Conventional Monte Carlo methods} \label{old_IKKT_numerical}
In this section, we review the numerical studies of the type IIB matrix model based on conventional Monte Carlo methods. In particular, we discuss simulations of the Euclidean IIB matrix model based on the factorization method, and simulations of the Lorentzian version based on an approximation to avoid the sign problem.

\subsection{Euclidean version of the type IIB matrix model} 
\label{review_EIKKT}

Historically, the Euclidean version of the type IIB matrix model and related simplified models were studied numerically \cite{9811220,0003208,0005147,0108041,1009_4504,1108_1534,1306_6135,1509_05079,1712_07562,2002_07410} before the Lorentzian IIB matrix model.
This is because
the Euclidean version is finite without any cutoff \cite{9803117,0103159}, and taking the large--$N$ limit is more straightforward. 
Furthermore,
the Euclidean model appears to be simpler to simulate because the action enters the partition function without an $i$, as in the conventional approach to simulating lattice quantum field theories. Therefore, the model without fermions is easy to simulate using the usual Monte Carlo methods, without a sign problem. Unfortunately, when we consider dynamical fermions, the system has a strong complex action problem, which has to be addressed with various methods.

In this section, we consider the factorization method~\cite{0108041,Ambjorn:2002pz,Azcoiti:2002vk,Fodor:2004nz,Ambjorn:2004jk,1009_4504,1306_6135,1509_05079}.
Using this approach, it is possible to obtain evidence that the SSB of SO(10) rotational symmetry to SO(3) occurs due to the strong fluctuations of the phase of the Pfaffian.

\subsubsection{The definition of the model and the SSB}
\label{EIKKT-SSB}
In the Euclidean version,
the temporal direction is Wick-rotated as 
\begin{align}
 A_{10} = -i A_{0}, \ \ \Gamma_{10} = i \Gamma^0 ,  \label{wick_rotation}
\end{align}
and the indices $\mu,\nu=1,2,\dots , 10$ 
are contracted using the Euclidean metric $\displaystyle \delta_{\mu \nu} = \textrm{diag} (1,1,1,\dots,1)$. We adopt the following representation for the gamma matrices:
\begin{align}
 \Gamma_1 &= i \sigma_2 \otimes \sigma_2 \otimes \sigma_2 \otimes \sigma_2, \ \ \Gamma_2 = i \sigma_2 \otimes \sigma_2 \otimes {\bf 1} \otimes \sigma_1, \ \  \Gamma_3 = i \sigma_2 \otimes \sigma_2 \otimes {\bf 1} \otimes \sigma_3 \nonumber \\
 \Gamma_4 &= i \sigma_2 \otimes \sigma_1 \otimes \sigma_2 \otimes {\bf 1}, \ \ \Gamma_5 = i \sigma_2 \otimes \sigma_3 \otimes \sigma_2 \otimes {\bf 1}, \ \  \Gamma_6 = i \sigma_2 \otimes {\bf 1} \otimes \sigma_1 \otimes \sigma_2 \nonumber \\
 \Gamma_7 &= i \sigma_2 \otimes {\bf 1} \otimes \sigma_3 \otimes \sigma_2, \ \ \Gamma_8 = i \sigma_1 \otimes {\bf 1} \otimes {\bf 1} \otimes {\bf 1}, \ \  \Gamma_9 = i \sigma_3 \otimes {\bf 1} \otimes {\bf 1} \otimes {\bf 1} \nonumber \\
 \Gamma_{10} &= {\bf 1} \otimes {\bf 1} \otimes {\bf 1} \otimes {\bf 1}, \label{gamma_euclidean} 
\end{align}
where $\sigma_i$ ($i=1,2,3$) are the Pauli matrices.

The partition function of the Euclidean model is given by
\begin{align}
 Z_{\textrm{E}} = \int dA d\psi e^{-(S_{\textrm{b}}+ S_{\textrm{f}})} 
= \int dA e^{-S_{\textrm{b}}} \textrm{Pf } {\cal M} = \int dA e^{-S_{\textrm{eff}}}, 
 \label{EIKKT_partition}
\end{align}
where ${\cal M}$ is a $16 (N^2-1) \times 16 (N^2-1)$ antisymmetric matrix, defined by the linear transformation
\begin{align}
  \psi_{\alpha} \to ({\cal M} \psi)_{\alpha} = (\Gamma^{\mu})_{\alpha \beta} [A_{\mu}, \psi_{\beta}] \, ,
  \label{calM_def}
\end{align}
acting on the linear space of traceless complex $N \times N$ matrices $\psi_{\alpha}$. The effective action $S_{\textrm{eff}}$ is defined as
\begin{align}
 S_{\textrm{eff}}= S_{\textrm{b}} - \log \textrm{Pf } {\cal M}. \label{s_eff}
\end{align}
The dynamical compactification of space-time occurs when the SO(10) rotational symmetry is spontaneously broken. 
As an order parameter for the SSB we use the ``moment of inertia tensor" defined by
\begin{align}
 T_{\mu \nu} = \frac{1}{N} \textrm{tr} (A_{\mu} A_{\nu}). \label{Tmunu}
\end{align}
We order its eigenvalues $\lambda_n$ ($n=1,2,\dots,10$) as $\lambda_{1} \geq \lambda_{2} \geq \dots \geq \lambda_{10}$ before taking the VEV. If $\langle \lambda_{1} \rangle, \dots, \langle \lambda_{d} \rangle$ grow and $\langle \lambda_{d+1} \rangle, \dots, \langle \lambda_{10} \rangle$ shrink in the large--$N$ limit, this signals the SSB of SO(10) to SO($d$); namely the dynamical compactification of the ten-dimensional space-time to $d$-dimensions. This scenario has been studied using the GEM in Refs.~\cite{1007_0883,1108_1293}.
The results for the SO$(d)$ symmetric vacua with $2 \leq d \leq 7$ are summarized as follows:
\begin{enumerate}
\item{The extent of the shrunken directions $r = \lim_{N\to \infty} \sqrt{\lambda_n}$ ($n=d+1, \dots,10$) is $r^2 \simeq 0.155$, which does not depend on $d$ (universal compactification scale).}
\item{The ten-dimensional volume of the Euclidean space-time does not depend on $d$ except for $d=2$ (constant volume property).
The volume is given by $V = (R_d)^d r^{10-d} = \ell^{10}$ with $l^2 \simeq 0.383$,
where we define the extent of the extended directions by $R_d = \lim_{N \to \infty} \sqrt{\lambda_n}$ ($n=1,2,\dots,d$).}
\item{The free energy takes a minimum value at $d=3$, which suggests the dynamical emergence of {\it three}-dimensional space-time.}
\end{enumerate}
While these are interesting dynamical properties, their relevance to our four-dimensional space-time remains unclear.

\subsubsection{Absence of the SSB without the complex phase}
The Euclidean model suffers from a severe complex action problem due to the complex phase of $\textrm{Pf } {\cal M}$. 
It is actually the strong fluctuations of this complex phase that provide the mechanism for the dynamical compactification of
space-time \cite{0003223,Nishimura:2000wf}.

Note first that
$\textrm{Pf } {\cal M}$ is real for nine-dimensional configurations $A_{10}=0$.
Moreover, $\textrm{Pf } {\cal M}$ vanishes for two-dimensional configurations
$A_{3}=\dots = A_{10}=0$.
Let us define the phase of the Pfaffian $\Gamma$ as 
\begin{align}
 \textrm{Pf } {\cal M} = \lvert \textrm{Pf } {\cal M} \rvert e^{i \Gamma}. \label{phase_def}
\end{align}
When a configuration is $d$-dimensional ($3 \leq d \leq 8$), we find that
\begin{align}
  \frac{\partial^m \Gamma}{\partial A_{\mu_1}^{a_1} \dots \partial A_{\mu_m}^{a_m}} = 0\, ,
  \label{phase0003223}
\end{align}
for $m=1,2,\dots,9-d$, where $A_{\mu}^a$ are the coefficients in the expansion $A_{\mu} = \sum_{a=1}^{N^2-1} A_{\mu}^a T^a$ with respect to the SU$(N)$ generators $T^a$. This is because the configuration is at most nine-dimensional, up to the $(9-d)$-th order of perturbations. Thus, the phase of $\textrm{Pf } {\cal M}$ becomes more stationary for lower dimensions for $3 \leq d \leq 8$. This suggests that the complex phase $\Gamma$ plays a key role in the dynamical compactification of space-time.

The four-dimensional version of the Euclidean type IIB matrix model was studied in Refs.~\cite{0003208,Ambjorn:2001xs} using Monte Carlo simulations. For this model $\det {\cal M}>0$, there is no sign problem, and the SO(4) rotational symmetry is not broken.

In Ref.~\cite{1509_05079}, the phase-quenched partition function
\begin{align}
  Z_{0} = \int d A e^{-S_{\textrm{b}}} \lvert  \textrm{Pf } {\cal M} \rvert =\int dA e^{-S_{\textrm{0}}}, \textrm{ where } S_{\textrm{0}} = S_{\textrm{b}} - \log \lvert  \textrm{Pf } {\cal M} \rvert
  \label{pq_partition}
\end{align}
was studied numerically.
In Fig.~\ref{factorization_EIKKT_result} (Left), the VEVs $\langle \lambda_n \rangle_0$ are plotted, where $\langle \cdots \rangle_0$ is the VEV with respect to the phase-quenched partition function (\ref{pq_partition}). We see that the $\langle \lambda_n \rangle_0$ converge to $l^2 \simeq 0.4$ at large $N$ for all $n=1,2,\dots,10$. These results are consistent with the absence of the  SSB of the SO(10) rotational symmetry, and the constant volume property with $l^2 \simeq 0.383$ predicted by the GEM.

\subsubsection{Monte Carlo studies using the factorization method}

The partition function (\ref{EIKKT_partition}) has a severe complex action problem, and it was studied using the factorization method 
in Ref.~\cite{1509_05079}. The partition functions  (\ref{EIKKT_partition}) and (\ref{pq_partition}) favor non overlapping regions of the
configuration space, and that leads to a serious overlap problem. Using the factorization method, we can overcome the overlap problem,
and perform importance sampling efficiently~\cite{0108041,1009_4504,1108_1534,1306_6135,1509_05079}.

The density of states $\rho_n(x)$ for the normalized observable ${\tilde \lambda}_n = \frac{\lambda_n}{\langle \lambda_n \rangle_0}$ has the factorization property
\begin{align}
 \rho_n (x) &= \left\langle \delta (x - {\tilde \lambda}_n ) \right\rangle = \frac{ \int dA e^{-S_{\textrm{0}}} e^{i \Gamma } \delta (x - {\tilde \lambda}_n )  }{\int dA e^{-S_{\textrm{0}}}} \left/  \frac{\int dA e^{-S_{\textrm{0}}} e^{i \Gamma} }{\int dA e^{-S_{\textrm{0}}} } \right. \nonumber \\
 &= \frac{1}{\langle e^{i \Gamma}\rangle_0}  \times \frac{\int dA e^{-S_{\textrm{0}}} \delta (x - {\tilde \lambda}_n ) }{\int dA e^{-S_{\textrm{0}}}  }   \times \frac{\int dA e^{-S_{\textrm{0}}} e^{i \Gamma } \delta (x - {\tilde \lambda}_n ) }{\int dA e^{-S_{\textrm{0}}} \delta (x - {\tilde \lambda}_n ) } \nonumber \\
 &= \frac{1}{\langle e^{i \Gamma}\rangle_0} \left\langle \delta (x - {\tilde \lambda}_n ) \right\rangle_0 \langle e^{i \Gamma} \rangle_{n,x} , \label{factorization_density}
\end{align}
where $\langle \cdots \rangle$, $\langle \cdots \rangle_0$ and $\langle \cdots \rangle_{n,x}$ are the VEVs with respect to the partition functions (\ref{EIKKT_partition}), (\ref{pq_partition}) and (\ref{constrained_action}), respectively, where
\begin{align}
 Z_{n,x} = \int dA e^{-S_{\textrm{0}}} \delta( x - {\tilde \lambda}_n ). \label{constrained_action}
\end{align}
We denote the quantities in Eq.~(\ref{factorization_density}) as 
\begin{align}
 C = \langle e^{i \Gamma}\rangle_0, \ \ \rho_n^{(0)} (x) = \left\langle \delta (x - {\tilde \lambda}_n ) \right\rangle_0, \ \ w_n (x) = \langle e^{i \Gamma} \rangle_{n,x}. \label{factorization_density2}
\end{align}
In this method, it is important  that there is no need to calculate $C$, and that we only have to calculate $w_n(x)$ in the relevant region of $x$.

By constraining the eigenvalue $\lambda_n$ to some value smaller than $\langle \lambda_n \rangle_0$ with $n=d+1$ \cite{1306_6135,1509_05079},
we can probe the SO($d$) vacuum, where the larger eigenvalues $\lambda_1,\dots, \lambda_{n-1}$ correspond to the extended directions.
When $N$ is large enough, the VEV 
is estimated by $\langle {\tilde \lambda}_n \rangle \simeq {\bar x}_n$, where ${\bar x}_n$ is the position of the peak of $\rho_n(x)$.
Thus, our task reduces to finding the minimum of the free energy,
\begin{align}
 {\cal F}_{\textrm{SO}(d)} (x) 
 = - \frac{1}{N^2} \log \rho_n (x) \quad 
\textrm{ with } n=d+1,
 \label{free_energy}
\end{align}
which amounts to solving the saddle-point equation 
\begin{align}
 \frac{1}{N^2} f_n^{(0)} (x) = - \frac{d}{dx} \frac{1}{N^2} \log w_n (x), \ \textrm{ where } f_n^{(0)} (x) = \frac{d}{dx} \log \rho_n^{(0)} (x). 
\label{saddlepoint_factorization}
\end{align}

\begin{figure} 
\centering
\vspace*{-5mm}
\includegraphics[width=0.32\textwidth]{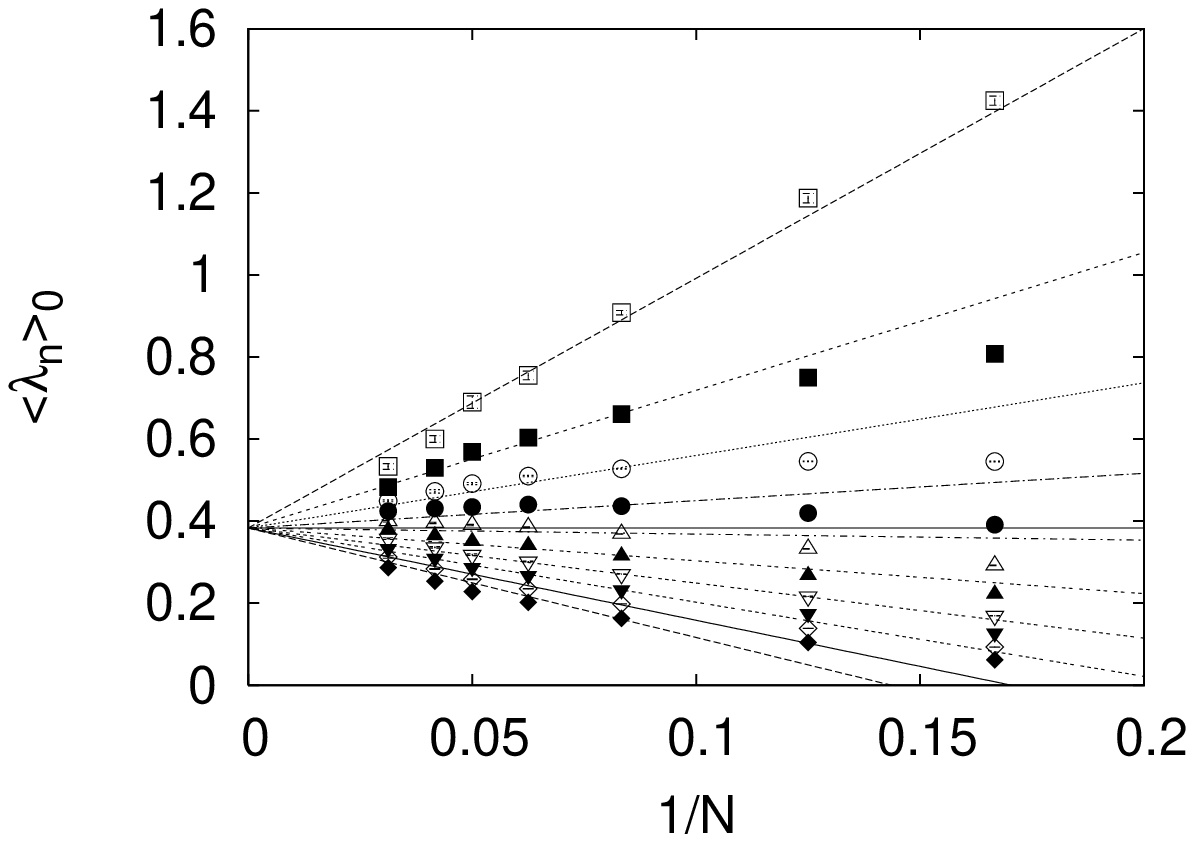}
\includegraphics[width=0.32\textwidth]{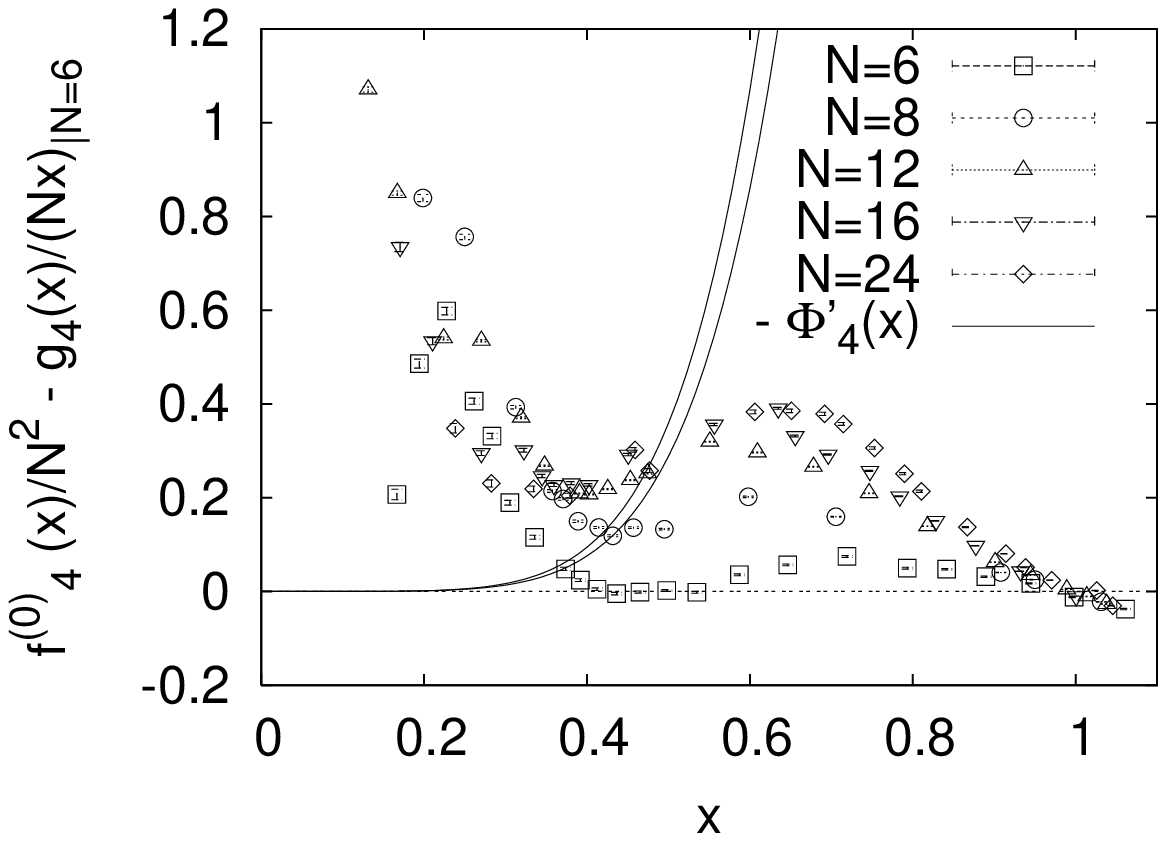}
\includegraphics[width=0.32\textwidth]{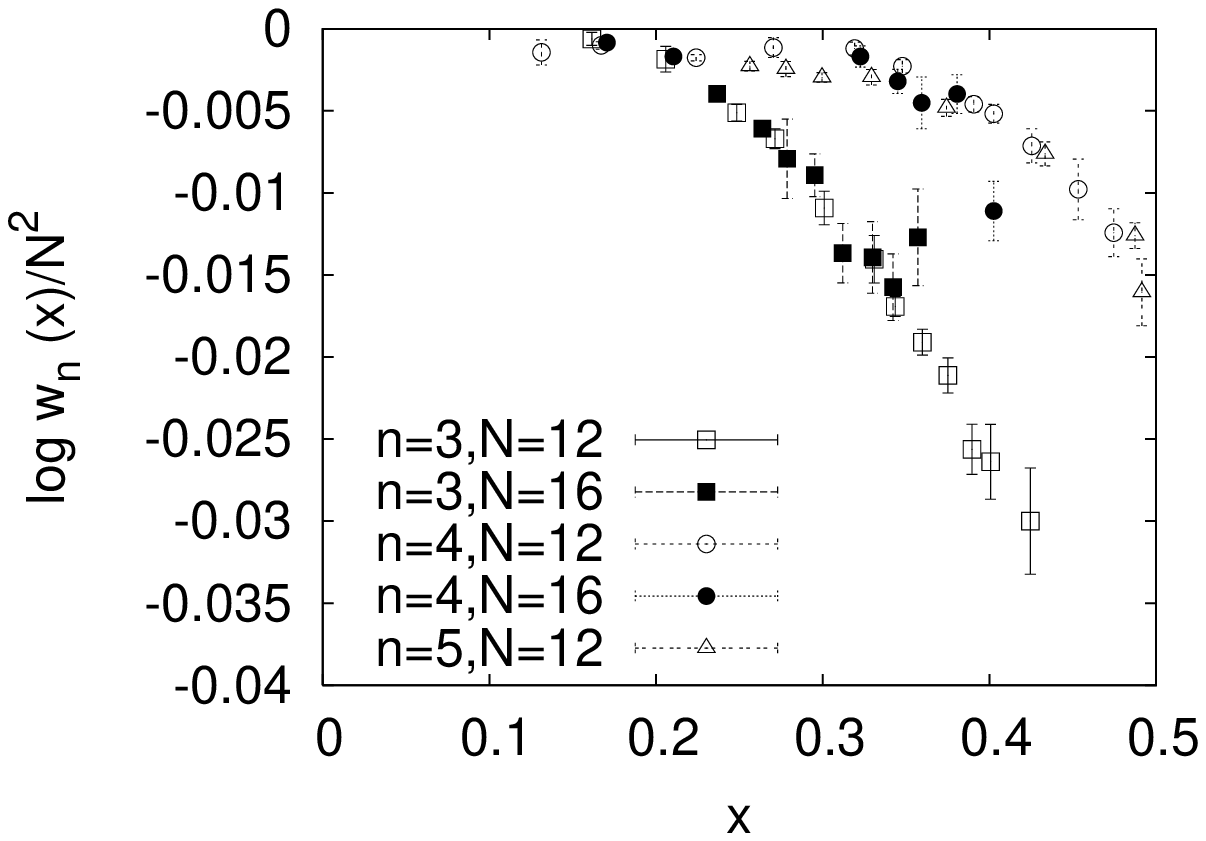}

\vspace*{5mm}
\caption{(Left) The VEVs $\langle \lambda_n \rangle_0$ with respect to the phase-quenched partition function (\ref{pq_partition}) up to $N=32$. Quoted from Fig.~1 of Ref.~\cite{1509_05079}. (Middle) $\frac{1}{N^2} f^{(0)}_n (x) - \frac{g_n (x)}{Nx}$ up to $N=24$, and  $- \frac{d}{dx} \Phi_n (x)$ for $n=4$. Their intersection gives the solution ${\bar x}_4$. Quoted from Fig.~2 of Ref.~\cite{1509_05079}. (Right) $\frac{1}{N^2} \log w_n (x)$ for $n=3,4$, $N=12,16$ and $n=5$, $N=12$. Quoted from Fig.~3 of Ref.~\cite{1509_05079}.}\label{factorization_EIKKT_result}
\end{figure}

To do this, we simulate the partition function (\ref{constrained_action}) using the RHMC (Rational Hybrid Monte Carlo) algorithm \cite{lat0309084},
whose details\footnote{In RHMC, we introduce the pseudofermion as in Eq.~(A.3) of Ref.~\cite{1306_6135}. Later, it turned out to be more convenient to update the pseudofermion by the heatbath algorithm instead of solving the Hamilton equation (A.6) of Ref.~\cite{1306_6135}.} are given in the Appendix A of Ref.~\cite{1306_6135}. The
quantity $w_n(x)$ has a large--$N$ scaling behavior at small $x$ given by
\begin{align}
 \frac{1}{N^2} \log w_n (x) \simeq - a_n x^{11-n} - b_n , \label{wn_scale}
\end{align}
which we use to obtain the large--$N$ limit
\begin{align}
 \Phi_n (x) = \lim_{N\to \infty} \frac{1}{N^2} \log w_n (x). \label{Phi_n_def}
\end{align}
The large--$N$ scaling behavior of  $f_n^{(0)} (x)$ in the region  $0.4 \leq x \leq 1$ is approximated by
\begin{align}
 \frac{x}{N} f^{(0)}_n (x) \simeq g_n (x), \textrm{ where }
  g_n (x) = c_n (x-1) + d_n (x-1)^2 . \label{fn_scaling}
\end{align}
In Fig.~\ref{factorization_EIKKT_result} (Middle), we plot  $\frac{1}{N^2} f^{(0)}_n (x) - \frac{g_n (x)}{Nx}$, where the second term is subtracted to 
reduce the finite-$N$ effects.
We find that the solution ${\bar x}_n$ for $n=4$ is close to the GEM prediction $\frac{r^2}{\ell^2} \simeq \frac{0.155}{0.383} = 0.404 \cdots$. 
Similar results are also obtained for $n=3$ and $n=5$.

The free energy at $x={\bar x}_n$ is given by
\begin{align}
 {\cal F}_{\textrm{SO}(d)} ({\bar x}_n)
 = \int^1_{{\bar x}_n} \frac{1}{N^2} f^{(0)}_n (x) dx - \frac{1}{N^2} \log w_n ({\bar x}_n) \quad \textrm{ with } n=d+1. \label{freeenergy_SOd}
\end{align}
The first term becomes negligible at large $N$ in the region $0.4 \leq x \leq 1$ due to the scaling \eqref{fn_scaling},
and  we can estimate Eq.~\eqref{freeenergy_SOd} from
$\frac{1}{N^2} \log w_n (x)$ at $\displaystyle x = {\bar x}_n \simeq 0.4$
obtained from Fig.~\ref{factorization_EIKKT_result} (Right). While it is difficult to
compare ${\cal F}_{\textrm{SO}(3)}$ and ${\cal F}_{\textrm{SO}(4)}$, we clearly see that ${\cal F}_{\textrm{SO}(2)}$ is higher than ${\cal F}_{\textrm{SO}(3)}$ and ${\cal F}_{\textrm{SO}(4)}$, which implies that the SO(2) vacuum
is disfavored. 

\subsection{Lorentzian version of the type IIB matrix model}
\label{review_LIKKT}

In this section, we review the results obtained  for the Lorentzian version of the type IIB matrix model 
in Refs.~\cite{1108_1540,1312_5415,1506_04795,1904_05914}. Here, we introduce infrared cutoffs
to cure the problems associated with the unbounded action and the flat directions
of the Lorentzian model. To avoid the sign problem due to the phase factor
$e^{iS_{\rm b}}$ in the partition function, we use an approximation, which enables us to investigate the model by conventional Monte Carlo methods.
We observe the emergence of (3+1)-dimensional expanding space-time, where the expansion is exponential at early times 
and power law at late times. On the other hand, it turns out that the space has a singular
Pauli-matrix structure, which implies that the emergent space is not smooth. It turns out that 
that the approximation used to avoid the sign problem  amounts to replacing $e^{iS_{\rm b}}$ by $e^{\beta S_{\rm b}}$ 
for  some positive coefficient $\beta$ (see Section \ref{CLM_sec}).

\subsubsection{Definition of the model}

In the Lorentzian model,
the indices are contracted using the Lorentzian metric $\displaystyle \eta_{\mu \nu} = \textrm{diag} (-1,1,1,\dots,1)$ and we {\it do not} 
Wick-rotate the matrices as in Eq.~(\ref{wick_rotation}). The gamma matrices are the same as those in Eq.~(\ref{gamma_euclidean}), except for
\begin{align}
 \Gamma_{0} = - \Gamma^0 = i \Gamma_{10} = i {\bf 1} \otimes {\bf 1} \otimes {\bf 1} \otimes {\bf 1}. 
\end{align}
The partition function for the Lorentzian model is given by
\begin{align}
 Z_{\textrm{L}} 
= \int dA d\psi e^{i (S_{\textrm{b}}+ S_{\textrm{f}})} 
= \int dA e^{iS_{\textrm{b}}} \textrm{Pf } {\cal M}. 
\label{LIKKT_partition}
\end{align}
\noindent 
The bosonic part of the action is not bounded from below because
\begin{align}
\textrm{tr} (F_{\mu \nu} F^{\mu \nu}) = -2 \textrm{tr} (F_{0I})^2 + \textrm{tr} (F_{IJ})^2, \label{fij_not_bounded}
\end{align}
where $\displaystyle F_{\mu \nu} = -i [A_{\mu}, A_{\nu}]$. To deal with this problem we introduce the infrared cutoffs $L$, $\kappa$ from 
the relations
\begin{align}
 &  \frac{1}{N} \textrm{tr} (A_0^2)^p \leq \kappa^p \frac{1}{N} \textrm{tr} (A_I^2)^p, \label{k_constraint} \\
 &  \frac{1}{N} \textrm{tr} (A_I^2)^p \leq L^{2p} \label{l_constraint}
\end{align}
for some $p$ ($1 \le p < 2$), as in Refs.~\cite{1108_1540,1312_5415,1506_04795,1904_05914}. 
While the Pfaffian $\textrm{Pf } {\cal M}$ is real in this case, the model suffers from the sign problem coming from the 
factor $e^{iS_{\textrm{b}}}$ in Eq.~\eqref{LIKKT_partition}.
To avoid this sign problem, we approximate the partition function (\ref{LIKKT_partition}) 
by\footnote{See the Appendix A of Ref.~\cite{1312_5415} for an argument for its justification.}
\begin{align}
 Z_{\textrm{L}}' &= \int dA \,  \textrm{Pf } {\cal M} \, \delta \left( \frac{1}{N} \textrm{tr} F_{\mu \nu} F^{\mu \nu}  \right) \delta \left( \frac{1}{N} \textrm{tr} \{(A_I)^2 \}^p - 1 \right) \nonumber \\
 &  \ \ \ \times \theta \left( \kappa^p - \frac{1}{N} \textrm{tr} \{ (A_0)^2  \}^p \right),  \label{LIKKT_partition_ap}
\end{align}
where $\theta(x)$ is the Heaviside step function. The partition function (\ref{LIKKT_partition_ap}), which is essentially free from the sign problem, 
has been studied using the RHMC algorithm. 

\subsubsection{The emergence of (3+1)D expanding space-time}

\begin{figure} 
\centering

\vspace*{-5mm}
\includegraphics[width=0.32\textwidth]{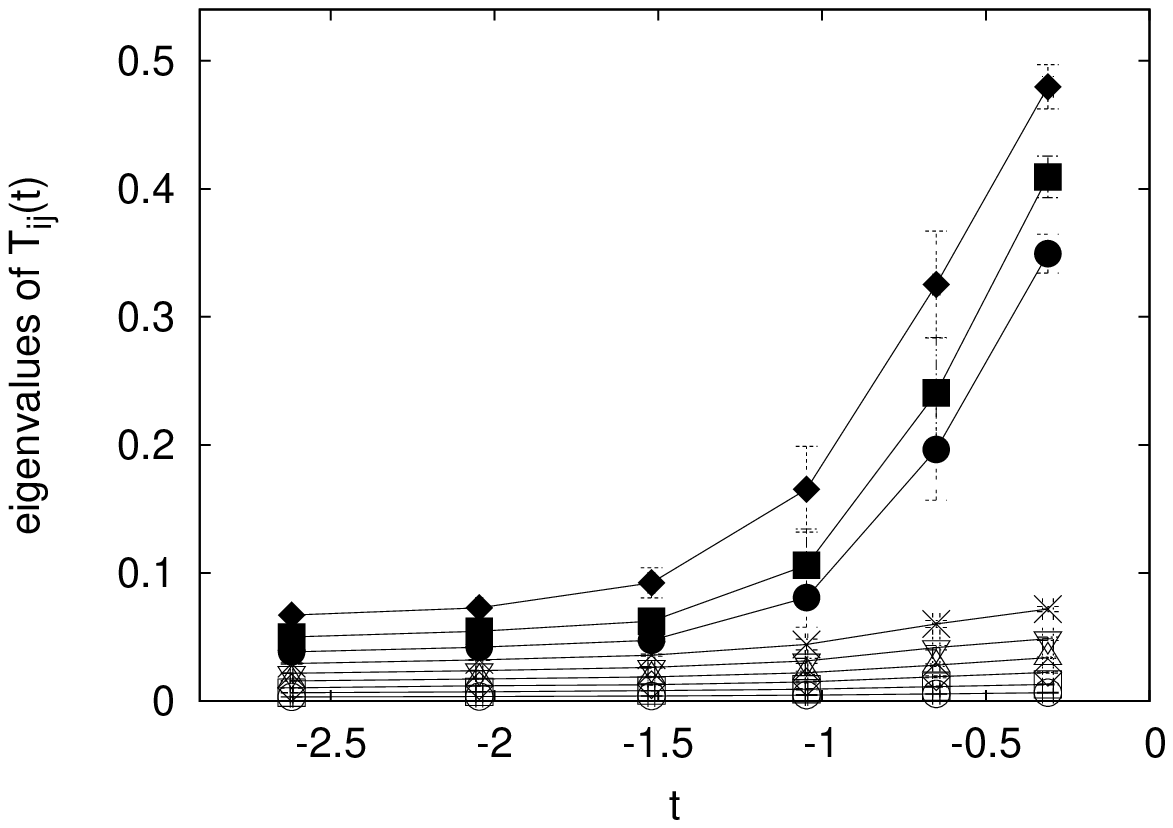}
\hspace*{0.5cm}
\includegraphics[width=0.32\textwidth]{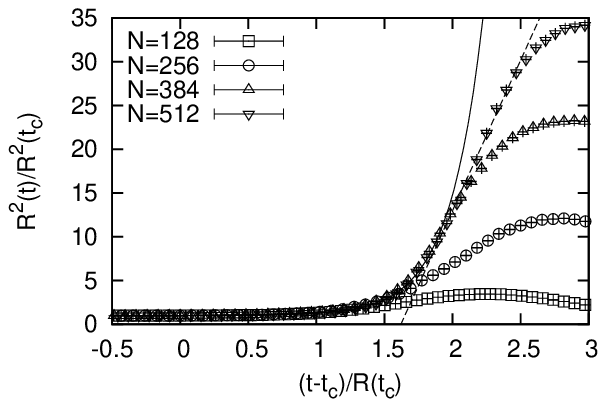}
\hspace*{-0.5cm}
\includegraphics[width=0.32\textwidth]{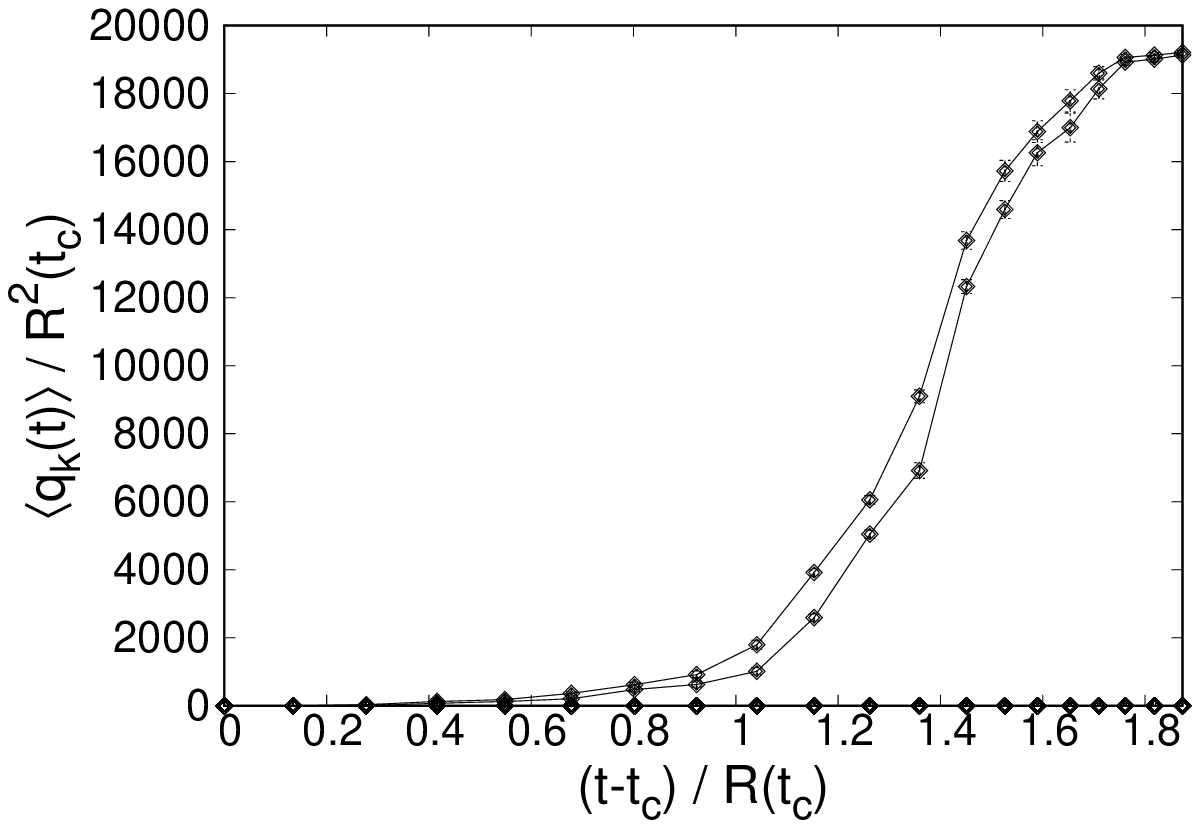}

\vspace*{8mm}
\caption{(Left) The 9 eigenvalues of $T_{IJ} (t)$ are plotted as a function of $t$ for $N=16, n=4, \kappa=4.0, L=1, p=1$. After the critical time $t_{\textrm{c}}$, 3 eigenvalues become larger. Quoted from Fig.~2 of Ref.~\cite{1108_1540}. (Middle) $R^2 (t)$ normalized by $R^2 (t_{\textrm{c}})$ are plotted against $x=(t-t_{\textrm{c}})/R(t_{\textrm{c}})$ for the bosonic model at $(N,n)=(128,20),(256,24),(384,28),(512,32),  L=1, p=1$, without the constraint (\ref{k_constraint}). Quoted from Fig.~6 of Ref.~\cite{1506_04795}. (Right) The eigenvalues of $Q(t)$ normalized by $R^2 (t_{\textrm{c}})$ are plotted against $x=(t-t_{\textrm{c}})/R(t_{\textrm{c}})$ for the bosonic model with $N=256, n=18, \kappa=1, L=1, p=1.5$. Quoted from Fig.~1 of Ref.~\cite{1904_05914}.}\label{sm1_0_result}
\end{figure}

To extract the time evolution of space from the matrix configurations, we fix the ``gauge'' by diagonalizing $A_0$ as
\begin{align}
 A_0 = \textrm{diag} (\alpha_1, \alpha_2, \dots, \alpha_N) 
\label{diagonal_gauge}
\end{align}
using an SU($N$) transformation. In this basis, the spatial matrices $A_i$ can have a band-diagonal structure with the bandwidth $n$.
When this happens in an actual simulation, we can regard the $n \times n$ submatrices
\begin{align}
 ({\bar A}_I)_{ab} (t_{\rho}) = (A_I)_{\rho+a, \rho+b} \ \ (a,b=1,2,\dots,n, \ \ \rho=0,1,2,\dots,N-n)  \label{space_for_t}
\end{align}
of the spatial matrices as a state corresponding to the time 
\begin{align}
 t_{\rho} = \frac{1}{n} \sum_{I=1}^n \alpha_{\rho+I}. \label{old_temporal_def}
\end{align}
We define the  $9 \times 9$ real symmetric tensor
\begin{align}
 T_{IJ} (t) = \frac{1}{n} \textrm{tr} \left( {\bar A}_I (t) {\bar A}_J (t) \right), \label{tensor_tij}
\end{align}
to be the order parameter of the SSB of SO(9) symmetry.
The 3 out of the 9 eigenvalues of the tensor $T_{IJ} (t)$ start to increase at a critical time $t_{\textrm{c}}$ as shown in Fig.~\ref{sm1_0_result} (Left), 
which suggests the SSB of SO(9) to SO(3)~\cite{1108_1540}.

It is interesting to study the expanding behavior of the universe in this model at late times \cite{1312_5415,1506_04795},
which requires simulations at large $N$. We define the extent of space at time $t$ as
\begin{align}
 R^2 (t) = \left\langle \frac{1}{N} \textrm{tr} \sum_{I=1}^9 ({\bar A}_I (t))^2 \right\rangle . \label{extent_r}
\end{align}
In Fig.~\ref{sm1_0_result} (Middle), $R^2 (t)$ normalized by $R^2 (t_{\textrm{c}})$ is plotted against $x=(t-t_{\textrm{c}})/R(t_{\textrm{c}})$ for 
the bosonic model. This suggests an exponential expansion of the universe at early times, followed by a power-law expansion at later times.

On the other hand,
the expanding three-dimensional space turns out to have a Pauli-matrix structure \cite{1904_05914}. To see it, let us introduce a $n \times n$ matrix
\begin{align}
 Q(t) = \sum_{I=1}^9 ({\bar A}_I(t))^2. \label{qq_def}
\end{align}
In Fig.~\ref{sm1_0_result} (Right), the eigenvalues $q_k (t)$ of $Q(t)$, normalized by $R^2 (t_{\textrm{c}})$, are plotted against 
$x=(t-t_{\textrm{c}})/R(t_{\textrm{c}})$ for the bosonic model \cite{1904_05914}. It was found that only 2 of the
eigenvalues expand\footnote{In Ref.~\cite{1904_05914}, the appearance of the Pauli-matrix structure was confirmed more explicitly. 
It was also shown that the situation does not change even in the large--$N$ limit or in the presence of supersymmetry.}.
This was attributed to the approximation to avoid the sign problem, which leads us to take the effect of the complex phase into full account.

\section{Complex Langevin Method}\label{CLM_sec}
In this section, we discuss how the type IIB matrix model can be studied by the CLM \cite{Parisi:1983mgm,Klauder:1983sp},
which is a promising way to study complex-action systems numerically. The first step in this direction was taken in Ref.~\cite{1904_05919}.
We review the results obtained by this method for the Euclidean and Lorentzian versions. In the Euclidean version,
we confirm the SSB from SO(10) to SO(3), which was suggested by the GEM as discussed in Section \ref{EIKKT-SSB}.
We also find that the Lorentzian version without any cutoff is actually equivalent to the Euclidean version,
which implies, in particular, that the space-time that emerges in the Lorentzian version is actually \emph{complex}.
This motivates us to introduce a Lorentz-invariant mass term to the action. When the coefficient is sufficiently large, 
a classical solution describing \emph{real} space-time with expanding behavior dominates the path integral.
Furthermore, preliminary results for the bosonic model suggest the possibility of the SSB of SO(9) symmetry, as 
the coefficient of the mass term gets smaller.

\subsection{Application of the CLM to the type IIB matrix model} \label{CLM_EIKKT_alg}
The CLM consists of solving the complexified version of the Langevin equation. The following exposition is based on the Euclidean version of the 
type IIB matrix model, but the Lorentzian version can be worked out similarly. The complex Langevin equation is given by 
\begin{align}
 \frac{d (A_{\mu})_{kl} (\sigma)}{d\sigma} = - \frac{\partial S_{\textrm{eff}}}{\partial (A_{\mu})_{lk}} + (\eta_{\mu})_{kl} (\sigma), \label{CLM_eq1}
\end{align}
where $S_{\textrm{eff}}$ is defined by Eq.~(\ref{s_eff}), and $\sigma$ is the fictitious Langevin time.
The traceless Hermitian matrices $(\eta_{\mu})_{kl} (\sigma)$ are stochastic variables which follow the probability distribution
$ \exp \left( - \frac{1}{4} \int d \sigma \textrm{tr} \eta^2_{\mu} (\sigma) \right)$. 
Note that the Hermiticity of $A_{\mu}$ cannot be maintained as the Langevin time $\sigma$ evolves
due to the complex action $S_{\textrm{eff}}$, which forces the $A_{\mu}$ to be general traceless complex matrices.

The VEV of an observable ${\cal O}$ is evaluated from
\begin{align}
 \langle {\cal O}[A_{\mu}] \rangle =\frac{1}{T} \int^{\sigma_0 + T}_{\sigma_0} d\sigma {\cal O} [A_{\mu} (\sigma)] , \label{VEV_obs}
\end{align}
where $\sigma_0$ is the thermalization time.
The holomorphicity of ${\cal O}$ plays an essential role in showing the validity of Eq.~(\ref{VEV_obs}) \cite{0912_3360,1101_3270,1606_07627}.

We call the term $\frac{\partial S_{\textrm{eff}}}{\partial (A_{\mu})_{lk}}$ in Eq.~\eqref{CLM_eq1} the drift term, which contains a term
\begin{align}
  - \frac{\partial }{\partial (A_{\mu})_{lk}} \log \textrm{Pf } {\cal M}
  = - \frac{1}{2} \textrm{Tr} \left( \frac{\partial {\cal M}}{\partial (A_{\mu})_{lk}} {\cal M}^{-1} \right), \label{pf_diff}
\end{align} 
where $\textrm{Tr}$ represents the trace with respect to a $16 (N^2-1) \times 16(N^2-1)$ matrix. 
The direct calculation of the trace has a cost of  O$(N^6)$ in CPU time, and we make use of the ``noisy estimator" \cite{Batrouni:1985jn} 
based on the identity
\begin{align}
 \textrm{Tr} \left( \frac{\partial {\cal M}}{\partial (A_{\mu})_{lk}}{\cal M}^{-1} \right) = \left\langle \chi^{*} \frac{\partial {\cal M}}{\partial (A_{\mu})_{lk}}{\cal M}^{-1} \chi \right\rangle_{\chi}. \label{noisy_estimator}
\end{align} 
The average $\left\langle \cdots \right\rangle_\chi$ is taken with respect to the Gaussian noise $\chi$, which represents a $16 (N^2-1)$-dimensional vector whose components are complex Gaussian random variables normalized as $\left\langle \chi_k^{*} \chi_l \right\rangle = \delta_{kl}$. The quantity $\zeta = {\cal M}^{-1} \chi$ is calculated by solving the linear equation 
$\displaystyle {\cal M}^{\dagger} {\cal M} \zeta = {\cal M}^{\dagger} \chi$ using the conjugate gradient (CG) method.
This is possible because  ${\cal M}^{\dagger} {\cal M}$ is positive semi-definite. 
The  multiplication of ${\cal M}$ and ${\cal M}^{\dag}$ is done  using Eq.~(\ref{calM_def}) 
and\footnote{Note that $A_\mu$ is no more Hermitian in the CLM.}
\begin{align}
 \psi_{\alpha} \to ({\cal M}^{\dag} \psi)_{\alpha} = (\Gamma_{\mu}^{\dag})_{\alpha \beta} [A_{\mu}^\dag , \psi_{\beta}],
  \label{calM_def-dag}
\end{align}
and each one costs O$(N^3)$ in CPU time. The total cost of the simulation is dominated by this procedure, and its scaling with $N$ depends on the number of CG steps necessary for convergence.

The Langevin equation (\ref{CLM_eq1}) is put on a computer by discretizing it as
\begin{align}
 (A_{\mu})_{kl} (\sigma +\Delta \sigma) = (A_{\mu})_{kl} (\sigma) - \Delta \sigma \frac{\partial S_{\textrm{eff}}}{\partial (A_{\mu})_{lk}} + \sqrt{\Delta \sigma} ({\tilde \eta}_{\mu})_{kl} (\sigma) , \label{RK2_a}
\end{align}
where $\Delta \sigma$ is the step size. 
The factor $\sqrt{\Delta \sigma}$ comes from the normalization of the noise ${\tilde \eta}_{\mu}$, which follows a probability distribution proportional 
to $\exp \left( - \frac{1}{4} \sum_{\sigma} \textrm{tr} {\tilde \eta}_{\mu}^2 (\sigma) \right)$. 

To extract a reliable result equivalent to the path integral from the CLM, we need to avoid the following two problems.
One is the excursion problem in the anti-Hermitian direction, which occurs when $A_{\mu}$ is too far from being Hermitian.
The other is the singular drift problem, which occurs when the drift term becomes large due to the eigenvalues of ${\cal M}$ 
accumulating near zero, as can be seen from Eq.~(\ref{pf_diff}). 

A useful criterion for the absence of
both these problems
has been proposed in Ref.~\cite{1606_07627}.
One computes the distribution of the drift norm
\begin{align}
 u = \sqrt{\frac{1}{10N^3} \sum_{\mu=1}^{10}\sum_{k,l=1}^{N} \left\lvert \frac{\partial S_{\textrm{eff}}}{\partial (A_{\mu})_{lk}} \right\rvert^2 }\, , \label{drift_norm}
\end{align}
and when it falls off exponentially or faster,
the CLM
can be justified.
This criterion is satisfied for all the parameter regions shown in this article.

In the Euclidean model, we do not introduce a ``gauge fixing'' (\ref{diagonal_gauge}), but in that case, we need to use the technique called gauge 
cooling~\cite{1211_3709,Nagata:2015uga} to keep $A_{\mu}$ close to Hermitian matrices. This technique amounts to making a complexified gauge transformation
after each step of solving the discretized Langevin equation in such a way that the Hermiticity norm
${\cal N}_{\textrm{H}} = - \frac{1}{10N} \sum_{\mu=1}^{10} \textrm{tr} (A_{\mu} - A_{\mu}^{\dag})^2$
is minimized.

\subsection{CLM for the Euclidean model} \label{review_EIKKT_CLM}

The CLM has been applied to the Euclidean model (\ref{EIKKT_partition}) in Refs.~\cite{1712_07562,2002_07410}.
In order to probe the SSB and avoid the singular drift problem, the action (\ref{s_eff}) was deformed by the terms~\cite{Ito:2016efb}
\begin{align}
 \Delta S_{\textrm{b}} &= \frac{N}{2} \varepsilon \sum_{\mu=1}^{10} m_{\mu} \textrm{tr} (A_{\mu})^2, \label{boson_mass} \\
 \Delta S_{\textrm{f}} &= - i m_{\textrm{f}} \frac{N}{2} \textrm{tr} (\psi_{\alpha} ({\cal C} \Gamma_8 \Gamma_9^{\dagger} \Gamma_{10})_{\alpha \beta} \psi_{\beta})\, ,
 \label{ferm_mass}
\end{align}
where the $m_{\mu}$ satisfy $0 < m_1 \leq \dots \leq m_{10}$.
We expect to see the SSB
by sending
the explicit rotational symmetry breaking parameter
to $\varepsilon = 0$
{\it after} taking the large--$N$ limit.
We consider the order parameter,
\begin{align}
  \lambda_{\mu} = \frac{1}{N} \textrm{tr} (A_{\mu})^2  \ \ \ (\mu=1,2,\dots,10),
  \label{SSB_CLM}
\end{align}
where there is no summation over $\mu$. This way, we avoid using a non-holomorphic observable, like the eigenvalues of the tensor (\ref{Tmunu}),
which would introduce subtleties in
obtaining correct expectation values by
the CLM. 
The mass term (\ref{ferm_mass}), on the other hand, is needed to avoid the singular drift problem in the CLM~\cite{Ito:2016efb}. 
Since it breaks the SO(10) symmetry to SO(7) $\times$ SO(3), we study whether the SO(7) 
symmetry is further broken to smaller groups as we decrease $m_{\textrm{f}}$. Notice that the original model is recovered after taking the 
limits $N\to\infty$, $\varepsilon\to 0$, $m_{\textrm{f}}\to 0$, in that order.

For $m_{\textrm{f}} \leq 1.4$ we take the $m_{\mu}$ to be
\begin{align}
  m_{\mu} = (0.5, 0.5, 1,2,4,8,8,8,8,8)\, ,
  \label{boson_mass_coef}
\end{align}
so that we can distinguish the SO$(d)$ vacua for $d=2,3,4,7$. First, we compute the ratio for finite $N$:
\begin{align}
 \rho_{\mu} (m_{\textrm{f}}, \varepsilon, N) =\frac{\langle \lambda_{\mu} \rangle_{m_{\textrm{f}},\varepsilon,N}}{\sum_{\nu=1}^{10} \langle \lambda_{\nu} \rangle_{m_{\textrm{f}},\varepsilon,N}}, \label{rho1}
\end{align}
where $\langle \lambda_{\mu} \rangle_{m_{\textrm{f}},\varepsilon,N}$ is the VEV with respect to the action deformed by adding the mass terms (\ref{boson_mass}) and (\ref{ferm_mass}) at finite $N$. Then, we make a large--$N$ extrapolation 
\begin{align}
  \rho_{\mu} (m_{\textrm{f}}, \varepsilon) = \lim_{N \to \infty} \rho_{\mu} (m_{\textrm{f}}, \varepsilon, N)
  \label{rho2}
\end{align}
based on the numerical results up to $N=128$. 
\begin{figure} 
\centering
\includegraphics[width=0.32\textwidth]{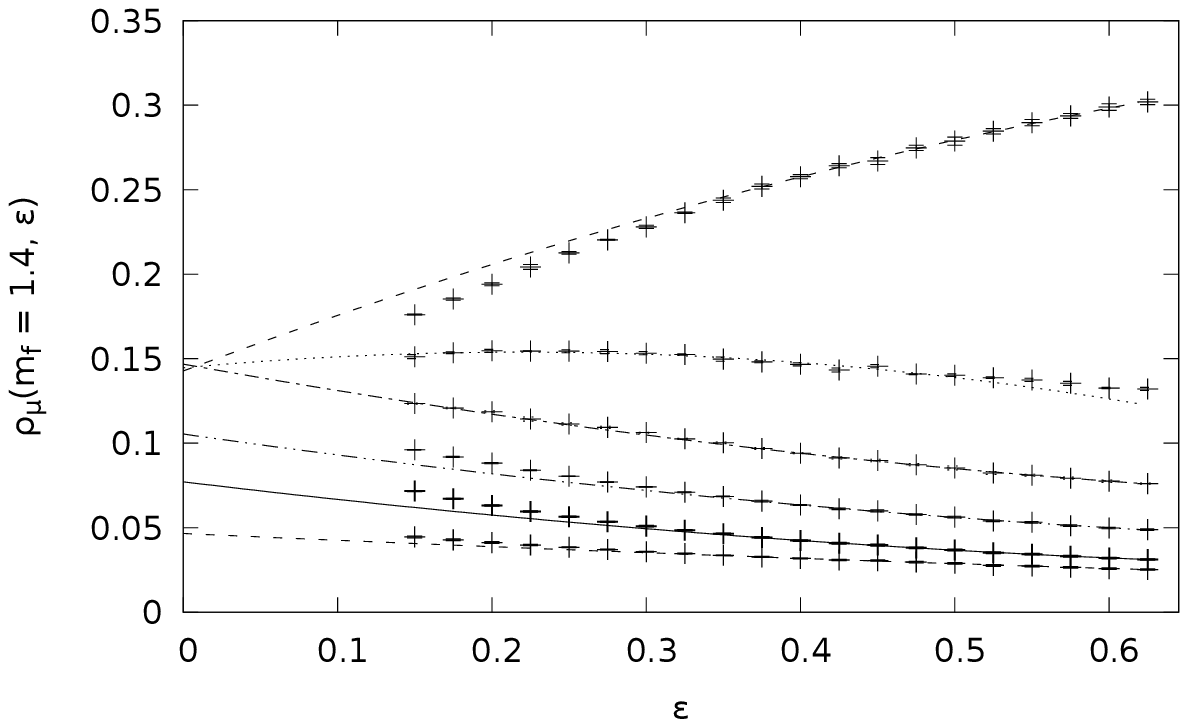}
\includegraphics[width=0.32\textwidth]{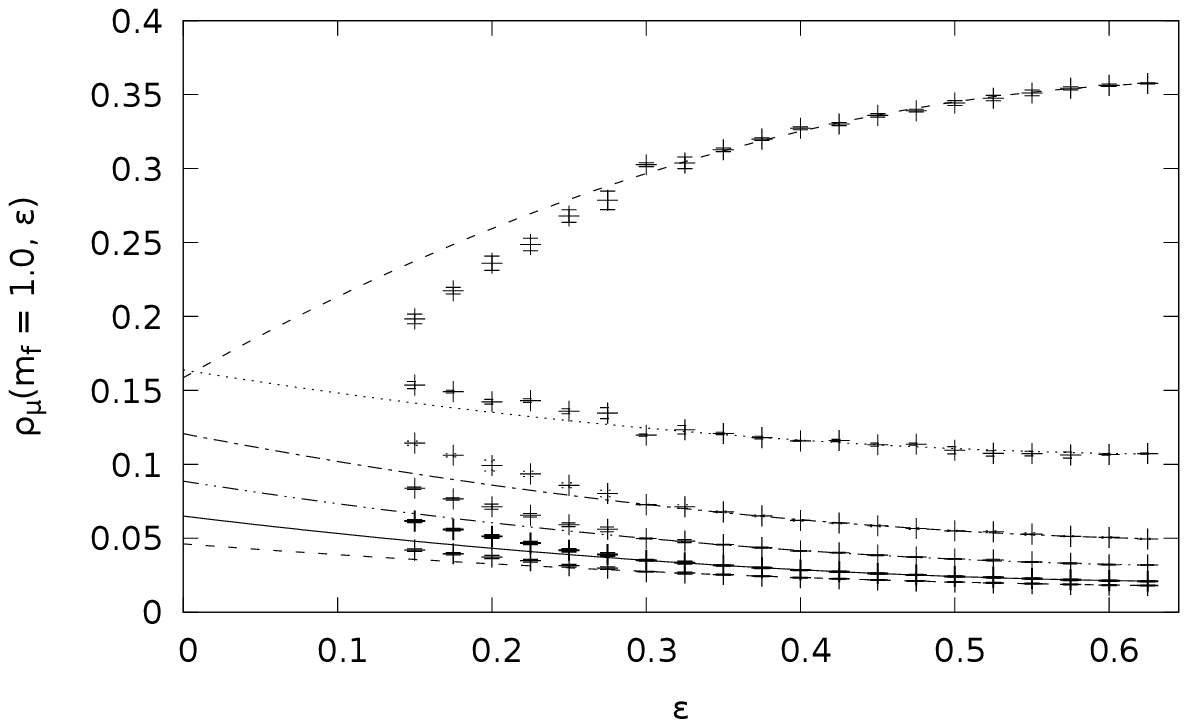}
\includegraphics[width=0.32\textwidth]{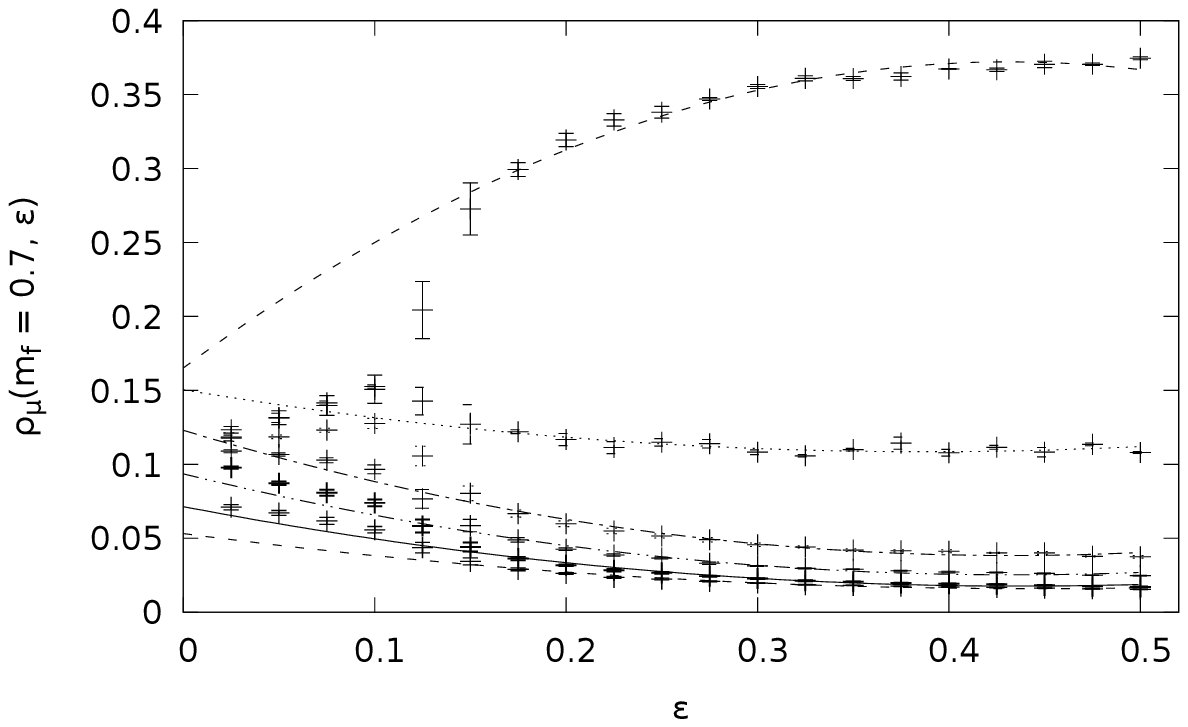}

\caption{$\rho_{\mu} (m_{\textrm{f}}, \varepsilon)$, which are obtained by the large--$N$ extrapolation based on $\rho_{\mu} (m_{\textrm{f}}, \varepsilon, N)$ up to $N=128$, are plotted for $m_{\textrm{f}}=1.4$ (Left), $m_{\textrm{f}}=1.0$ (Middle) and $m_{\textrm{f}}=0.7$ (Right). Quadratic fits are made with respect to $\varepsilon$. The curves from top to bottom are $\frac{1}{2} (\rho_1+\rho_2)$, \ $\rho_3$, \ $\rho_4$, \ $\rho_5$, $\frac{1}{2} (\rho_6+ \rho_7)$ and $\frac{1}{3} (\rho_8 + \rho_9 + \rho_{10})$, where we take averages for $\mu=1,2$, $\mu=6,7$ and $\mu=8,9,10$ to increase statistics. Quoted from Fig.~2 of Ref.~\cite{2002_07410}.}\label{CLM_EIKKT_result}
\end{figure}
In Fig.~\ref{CLM_EIKKT_result}, we plot $\rho_{\mu} (m_{\textrm{f}}, \varepsilon)$ against $\varepsilon$. Some of the small-$\varepsilon$ data are excluded from the fitting, considering the difficulty in large--$N$ extrapolations in the presence of SSB. We see that the SO(7) symmetry is broken to SO(4) for $m_{\textrm{f}} = 1.4$, and to SO(3) for $m_{\textrm{f}} = 0.7, 1.0$. This nicely confirms the GEM prediction for the type IIB matrix model ($m_{\rm f}=0$)
that the SO(10) symmetry is broken to SO(3).

\subsection{CLM for the Lorentzian model} \label{CLM_LIKKT_sec}
In this section, we apply the CLM to the Lorentzian version of the type IIB matrix model, which suffers from the sign problem due to the $e^{i S_{\textrm{b}}}$
in the partition function \eqref{LIKKT_partition}. 
For simplicity, we consider the bosonic model, which omits $\textrm{Pf } {\cal M}$.

The time evolution is extracted similarly to the way presented in Sec.~\ref{review_LIKKT}. In order to respect the ascending order $\displaystyle \alpha_1 \leq \alpha_2 \leq \cdots \leq \alpha_N$ when we complexify $\alpha_k$ in the CLM, we introduce the variables $\tau_k$ ($k=1,2,\dots,N-1$) as \cite{1904_05919}
\begin{align}
 A_0 = \textrm{diag} (\alpha_1, \dots, \alpha_N) \quad \textrm{ with } \alpha_1 = 0, \ \ \alpha_i =\sum_{k=1}^{i-1} e^{\tau_k} \ \ (i=2,3,\dots, N). \label{tau_def}
\end{align}
We also adopt the following definition of time, instead of Eq.~(\ref{old_temporal_def}):
\begin{align}
 t_{0} = 0 , \ \ t_{\rho} = \sum_{k=1}^{\rho} \lvert \bar{\alpha}_{k+1} - \bar{\alpha}_k \rvert,  \textrm{ where } \label{time_def}  \ \
 {\bar \alpha}_{k+1} = \frac{1}{n} \sum_{i=1}^{n} \alpha_{k+i}, 
\end{align}
with $\rho=1,2,\dots, N-n$ and $k=0,1,\dots,N-n$.

Let us introduce the parameters $s,k$ in the action as follows \cite{1904_05919}.
\begin{enumerate}
\item{We multiply an overall factor $e^{\frac{i}{2} s \pi}$, which corresponds to 
the Wick rotation on the worldsheet.}
\item{We make a replacement $A_0 \to A_{0} e^{-  \frac{i}{2} k \pi}$, 
which corresponds to the Wick rotation in the target space.}
\end{enumerate}
Then the partition function (\ref{LIKKT_partition}) becomes\footnote{If we do not omit the Pfaffian,
  it should be replaced as $\displaystyle \textrm{Pf } {\cal M} (A_0, A_I) \to \textrm{Pf } {\cal M} (A_0 e^{-  \frac{i}{2} k \pi}, A_I)$ considering that
  the overall phase of the fermionic action is irrelevant.}
\begin{align}
 Z &= \int d A \, e^{-{S}^{(s,k)}}, 
\label{LIKKT_partition2} \\
 {S}^{(s,k)} &= N  \left( - \frac{1}{2} e^{\frac{i}{2} (1+s-2k) \pi} \textrm{tr} [A_0,A_I]^2 - \frac{1}{4} e^{\frac{i}{2} (s-1) \pi} \textrm{tr} [A_I,A_J]^2 \right) ,
\label{Seff_sk}
\end{align}
where the indices $I,J$ run over $1,2,\dots,9$. The $(s,k)=(0,0)$ case is the Lorentzian model, which we are finally interested in. The $(s,k)=(1,1)$ case is the Euclidean model, which we reviewed in Sec.~\ref{review_EIKKT} and \ref{review_EIKKT_CLM}. The $(s,k)=(-1,0)$ case corresponds to the approximated partition function (\ref{LIKKT_partition_ap}), which we reviewed in Sec.~\ref{review_LIKKT}.

\begin{figure} 
\centering
\includegraphics[width=0.27\textwidth]{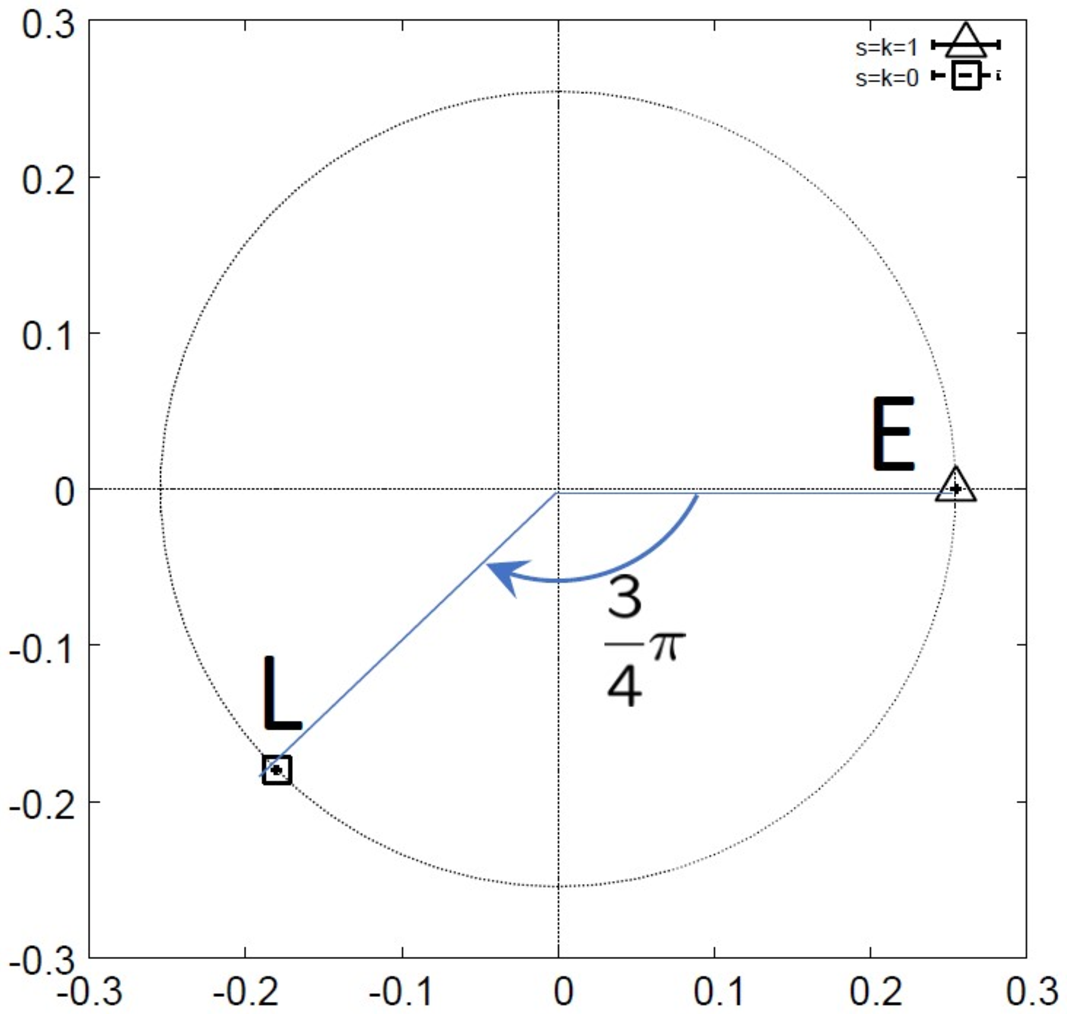}
\includegraphics[width=0.27\textwidth]{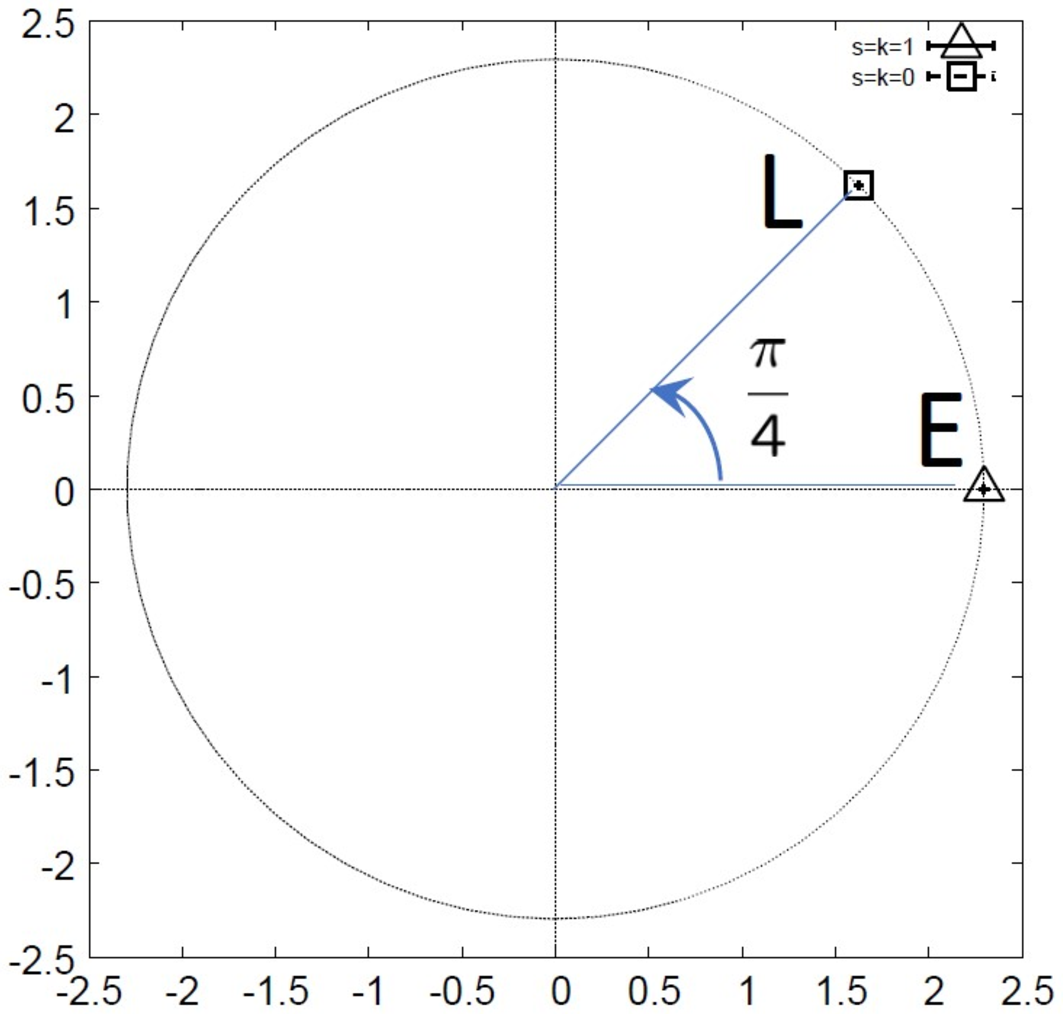}

\caption{VEVs of $\frac{1}{N} \textrm{tr} (A_0)^2$ (Left) and $\frac{1}{N} \textrm{tr} (A_0)^2$ (Right) for the $N=32$ bosonic case. E and L denote the VEVs in the Euclidean and Lorentzian models, respectively. Quoted from Fig.~1 of Ref.~\cite{2201_13200}.}\label{rel_EL_plot}
\end{figure}

Note that
the action satisfies
\begin{align}
{S}^{(0,0)} [A_0, A_I] = {S}^{(1,1)} [ {\tilde A}_0, {\tilde A}_I ], \label{rel_EL2}
\end{align}
where
\begin{align}
  A_0 = e^{-i\frac{3\pi}{8}} {\tilde A}_0, \ \ A_I = e^{i \frac{\pi}{8}} {\tilde A}_I .
  \label{rel_EL1}
\end{align}
This implies that, if the Lorentzian model is defined by the contour deformation \eqref{rel_EL1}
from the Euclidean model\footnote{This seems to be done automatically   when one applies the CLM to the Lorentzian model without cutoffs
  as one can see from Fig.~\ref{rel_EL_plot}.}, then the two models are equivalent due to Cauchy's theorem.
The following relation has been confirmed numerically in Fig.~\ref{rel_EL_plot} \cite{2201_13200}:
\begin{align}
 \left\langle \frac{1}{N} \textrm{tr} (A_0)^2 \right\rangle_{\textrm{L}} = e^{-i\frac{3\pi}{4}} \left\langle \frac{1}{N} \textrm{tr} ({\tilde A}_0)^2 \right\rangle_{\textrm{E}}, \ \ \left\langle \frac{1}{N} \textrm{tr} (A_I)^2 \right\rangle_{\textrm{L}} = e^{i\frac{\pi}{4}} \left\langle \frac{1}{N} \textrm{tr} ({\tilde A}_I)^2 \right\rangle_{\textrm{E}},  \label{rel_EL3}
\end{align}
where the VEVs $\langle \cdots \rangle_{\textrm{L}}$ and $\langle \cdots \rangle_{\textrm{E}}$ correspond to the Lorentzian model ${S}^{(0,0)}$ and the Euclidean model ${S}^{(1,1)}$, respectively. We see that neither time nor space is real in the Lorentzian model, and that the emergent space-time is interpreted to be Euclidean.

This observation leads us to add the Lorentz-invariant mass term 
\begin{align}
S_{\gamma} &= 
-  \frac{1}{2} N \gamma \textrm{tr} (A^{\mu} A_{\mu}) = \frac{1}{2} N \gamma \{ \textrm{tr} (A_0)^2 - \textrm{tr} (A_I)^2 \}
\end{align}
to the action \eqref{IKKT_action} \cite{2201_13200,2205_04726}. The coefficient $\gamma$ of the mass term 
is a positive parameter and it is sent to zero after taking the large--$N$ limit. One can view this mass term as an infrared regulator for the Lorentzian model,
as it will become clearer in the presence of $\varepsilon$ in Eq.~\eqref{massive_LIKKT2}. In this regard, note also that the redefinition \eqref{rel_EL1}
leads to the corresponding mass term for the Euclidean model, which is actually unbounded from below\footnote{This is not the case if $\gamma<0$.}.
Therefore, the mass term is expected to have a significant impact on the equivalence to the Euclidean model, which otherwise appears to be robust.

In fact,
the same mass term was introduced earlier in the study of classical solutions of the Lorentzian type IIB matrix model \cite{Hatakeyama:2019jyw}.
The classical equation of motion with the mass term is given by,
\begin{align}
\label{classical_eom}
[A^\nu, [A_\nu, A_\mu]]-\gamma A_\mu = 0.
\end{align}
For $\gamma>0$, the classical solutions with smooth and expanding space are obtained.
For $\gamma=0$, the solutions are simultaneously diagonalizable $A_\mu$,
which may or may not have expanding behavior. For $\gamma<0$, there are no classical solutions with expanding behavior.

The partition function of the model that we study using the CLM is given by
\begin{align}
  Z &= \int dA e^{-S_{\textrm{eff}}},
  \label{massive_LIKKT}  \\
 S_{\textrm{eff}} &= - i N  \left( \frac{1}{2} \textrm{tr} [A_0,A_I]^2 - \frac{1}{4} \textrm{tr} [A_I,A_J]^2 \right) - \frac{i}{2} N \gamma \left\{ e^{i \varepsilon} \textrm{tr} (A_0)^2 - e^{-i \varepsilon} \textrm{tr} (A_I)^2 \right\} \nonumber \\
 &- \log \prod_{1 \leq k<l \leq N} (\alpha_k- \alpha_l)^2 - \sum_{k=1}^{N-1} \tau_k , \label{massive_LIKKT2} 
\end{align}
where $\varepsilon>0$ in the mass term is introduced to make the integral absolutely convergent\footnote{The CLM works even at $\varepsilon=0$ and the 
results agree with the $\varepsilon \rightarrow +0$ limit.}. The two terms in the second line of Eq.~\eqref{massive_LIKKT2} represent
the Fadeev-Popov determinant associated with the gauge fixing \eqref{diagonal_gauge}
and the Jacobian associated with the change of variables (\ref{tau_def}), respectively.
We solve the complex Langevin equation in a manner similar to that presented in Sec.~\ref{CLM_EIKKT_alg} 
\begin{align}
 \frac{d (A_I)_{kl}}{d \sigma} = - \frac{\partial S_{\textrm{eff}}}{\partial (A_I)_{lk}} + (\eta_I)_{kl} (\sigma), \ \ \ \frac{d \tau_{a}}{d \sigma} = - \frac{\partial S_{\textrm{eff}}}{\partial \tau_a} + \eta_a (\sigma), \label{CL_eq_LIKKT}
\end{align}
where $(\eta_I)_{kl} (\sigma)$ and $\eta_a (\sigma)$ are Hermitian matrices
and real numbers respectively, following a probability distribution proportional to 
$\exp \left( - \frac{1}{4} \int d \sigma \textrm{tr} \eta_I^2 (\sigma) \right)$ 
and $\exp \left( - \frac{1}{4} \int d \sigma  \eta_a^2 (\sigma) \right)$, respectively.
The Langevin equations (\ref{CL_eq_LIKKT}) are discretized using the second-order Runge--Kutta method \cite{Fukugita:1986tg}
\begin{align}
 & (A_{I})_{kl} (\sigma +\Delta \sigma) = (A_{I})_{kl} (\sigma) + \sqrt{\Delta \sigma} ({\tilde \eta}_{I})_{kl} (\sigma) \nonumber \\
 & \ \ \ \ \ \ - \Delta \sigma \Biggl\{ \beta_1 \left[ \frac{\partial S_{\textrm{eff}}}{\partial (A_I)_{lk}}  (A(\sigma)) \right]+ \beta_2 \left[ \frac{\partial S_{\textrm{eff}}}{\partial (A_I)_{lk}}  (A'(\sigma)) \right] \Biggr\} , 
\label{RK2_a1} \\
 & (A'_{I})_{kl} (\sigma) = (A_{I})_{kl} (\sigma)  + \sqrt{\Delta \sigma} ({\tilde \eta}_{I})_{kl} (\sigma) - \Delta \sigma \left[ \frac{\partial S_{\textrm{eff}}}{\partial (A_I)_{lk}}  (A(\sigma)) \right] , \label{RK2_a2}
\end{align}
where
$\Delta \sigma$ is the step size, and the coefficients are taken to be $ \beta_1 =\beta_2 = \frac{1}{2} \left( 1 + \frac{N}{6} \Delta \sigma \right)$. The factor $\sqrt{\Delta \sigma}$ comes from the normalization of the noise ${\tilde \eta}$,
which follows the probability distribution $\propto \exp \left( - \frac{1}{4} \sum_{\sigma} \textrm{tr} {\tilde \eta}_I^2 (\sigma) \right)$. Note that
the same ${\tilde \eta}_I (\sigma)$ is used in Eqs. (\ref{RK2_a1}) and (\ref{RK2_a2}).
The update of $\tau_a$ is done similarly. 

We show our results for various values of $\gamma$ for $N=32$ and $\varepsilon=0$. We find that the simulation becomes unstable for large $\gamma$.
To stabilize the simulation, we insert a procedure
\begin{align}
A_I \mapsto \frac{A_I + \eta A_I^\dagger}{1+\eta} \quad \text{for } I=1,\dots,9
\end{align}
after each Langevin step, where the real positive parameter $\eta$ should be taken as small as possible\footnote{This is similar to the 
dynamical stabilization used in Lattice QCD applications of the CLM; its justification is not rigorous~\protect\cite{Attanasio:2018rtq}.}. 
Note that 
$\eta=1$ corresponds to Hermitianizing $A_I$ and $\eta=0$ corresponds to doing nothing. In the following, we set $\eta$ to $0.01$.
First, we obtain results for $\gamma=7$, and then decrease $\gamma$ adiabatically to obtain results for smaller $\gamma$.
The results at $\gamma=1$ are almost the same as those at $\gamma=0$, and in fact we can take $\eta=0$ for $\gamma=1$.

\begin{figure}%
\centering
\includegraphics[width=0.32\textwidth]{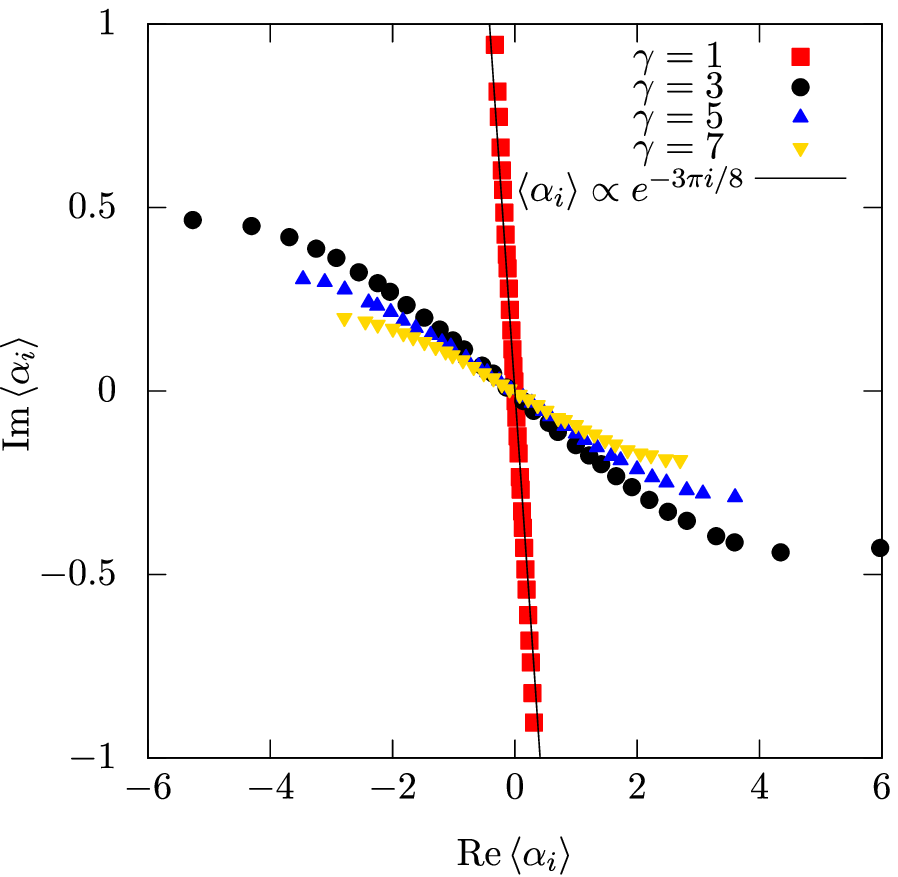}
\includegraphics[width=0.32\textwidth]{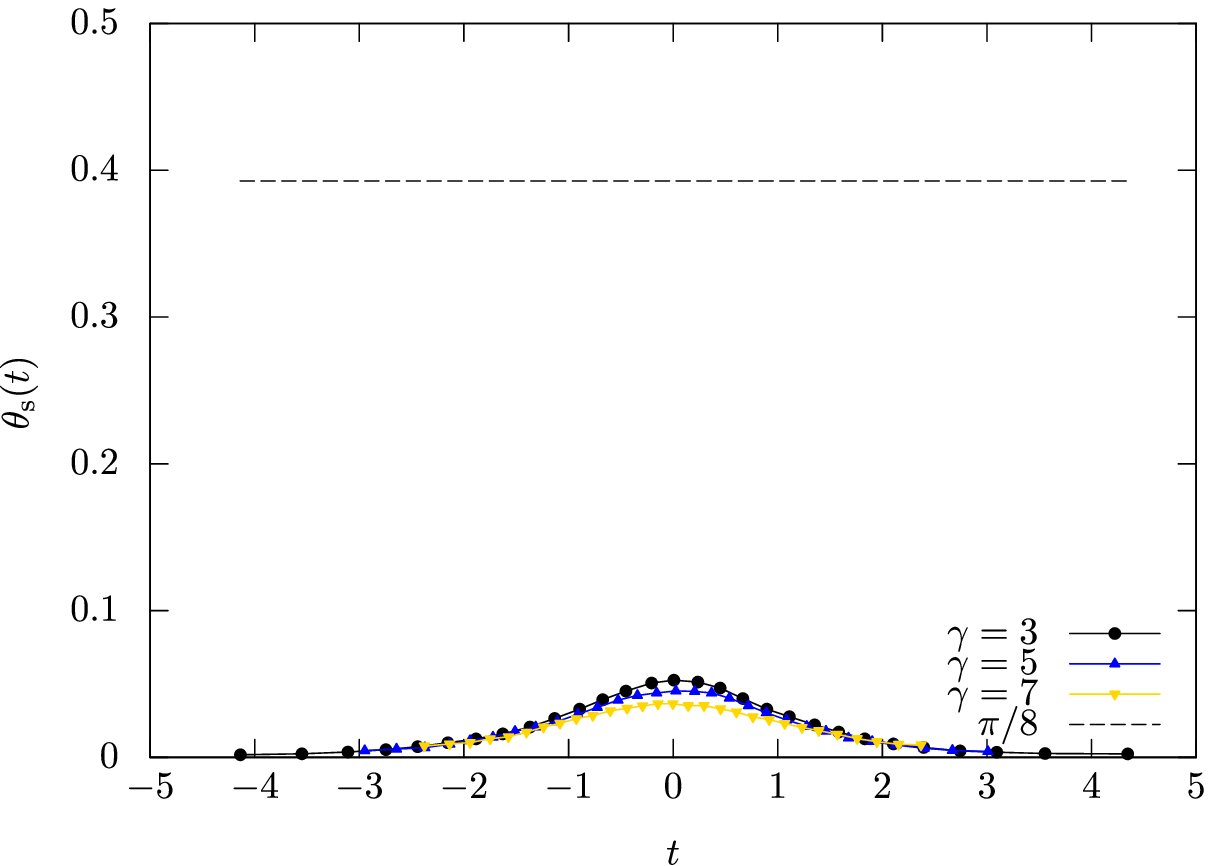}
\includegraphics[width=0.32\textwidth]{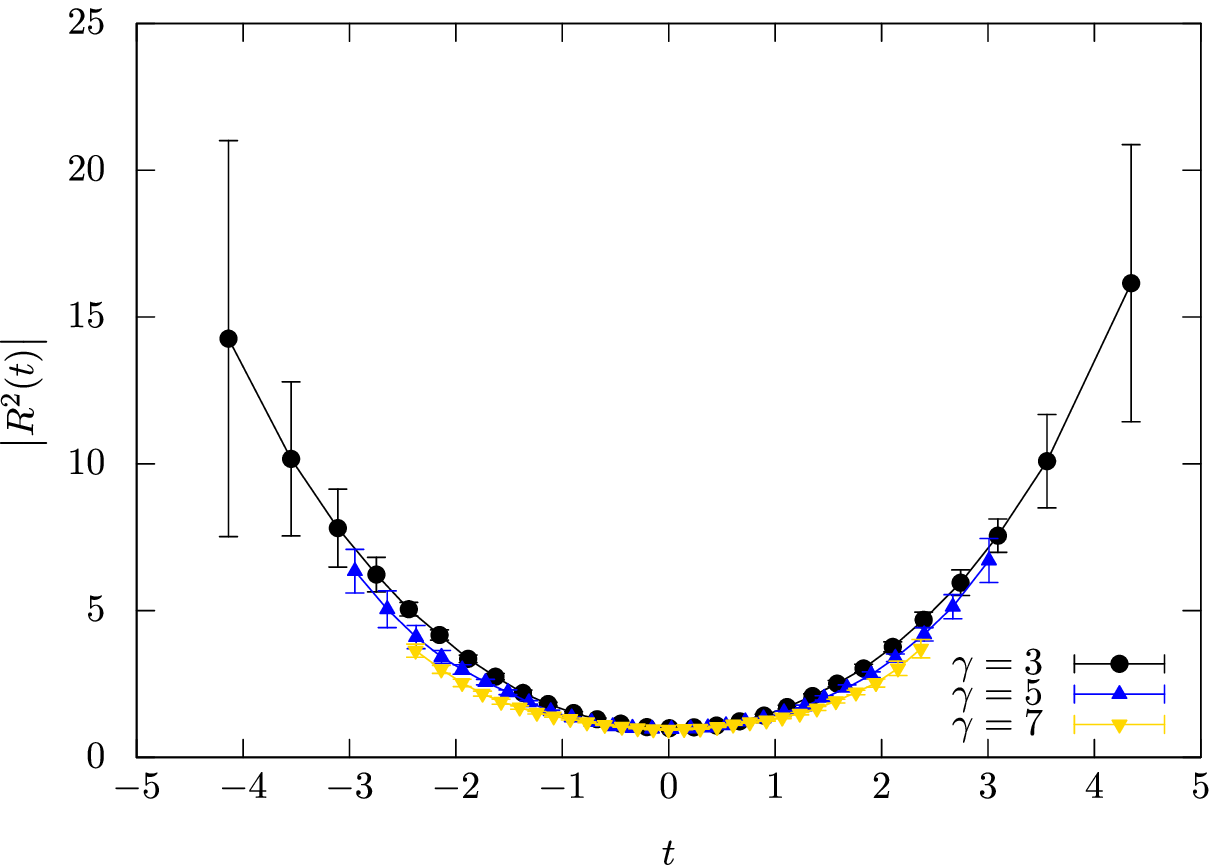}

\caption{(Left) The expectation values of $\alpha_i$ are plotted in the complex plane for $\gamma=1,3,5,7$. The solid line corresponds to the Euclidean model (see Fig.~\ref{rel_EL_plot} (Left)).
(Middle) The complex phase of $R^2(t)$, $\theta_\mathrm{s}(t)$, is plotted against the time $t$ for $\gamma=3,5,7$. The dashed line $\theta_\mathrm{s}(t)=\pi/8$ corresponds to the Euclidean model (see Fig.~\ref{rel_EL_plot} (Right)).
(Right) The extent of space $\vert R^2(t) \vert$ is plotted against the time $t$ for $\gamma=3,5,7$.
Quoted from Figs.~1 and 2 of Ref.~\cite{2205_04726}.\label{real_space-time}}
\end{figure}

In Fig.~\ref{real_space-time} (Left), we plot $\langle \alpha_i \rangle$ in the complex plane for $\gamma=1,3,5,7$.
Notice the chosen aspect ratio $1:6$.
For $\gamma=7$, the distribution of $\langle \alpha_i \rangle$ is close to the real axis, which suggests
that one of the classical solutions represented by Hermitian $A_\mu$
dominates\footnote{This can be understood theoretically by rescaling the matrices as $A_\mu \mapsto \sqrt{\gamma} \tilde{A}_\mu$, which brings the partition function into the form $Z \sim \int d\tilde{A} e^{- \gamma^2 \tilde{S}[\tilde{A}]}$. Since $\gamma$ appears only in the exponent, the path integral is dominated by some saddle-point configuration at large $\gamma$.} at large $\gamma$. 
As $\gamma$ becomes smaller, we observe that the distribution moves away from the real axis, but the flat region at both ends extends, suggesting the emergence of real time in that region\footnote{This property can be understood as a result of classicalization due to the expanding space, which makes the action large
  at late times \cite{Kim:2011ts,Kim:2012mw}.}. 
The result for $\gamma=1$ is close to the Euclidean model, and it is qualitatively different from the results for $\gamma \geq 3$.

In order to see the properties of space, we consider
\begin{align}
R^2(t) = e^{2i \theta_\mathrm{s}(t)} \vert R^2(t) \vert\ .
\end{align}
Note that $R^2(t)$ defined in Eq.~\eqref{extent_r} is complex due to the complex weight in the partition function 
for the Lorentzian model \eqref{LIKKT_partition}. In particular, $\theta_\mathrm{s}(t)=0$ corresponds to real space, 
while $\theta_\mathrm{s}(t)=\pi/8$ corresponds to the Euclidean model. In Fig.~\ref{real_space-time} (Middle) and (Right),
we plot $\theta_\mathrm{s}(t)$ and $\vert R^2(t) \vert$, respectively, against the time $t$ for $\gamma=3,5,7$.
The block size used in defining $\bar{\alpha}_{k}$ in Eq.~\eqref{time_def} and $\bar{A}_I(t)$ in Eq.~\eqref{space_for_t} is chosen to be $n=4$. 
From Fig.~\ref{real_space-time} (Middle), we find that space becomes real at late times, while the phase $\theta_\mathrm{s}(t)$ becomes slightly positive near $t \sim 0$. 
From Fig.~\ref{real_space-time} (Right), we find that the expanding behavior of $\vert R^2(t) \vert$ is analogous to that observed for classical solutions \cite{Hatakeyama:2019jyw}.
Scaling behavior is observed for different values of $\gamma$, and decreasing $\gamma$ results in extending the time direction,
and making space more expanded at late times. 

\begin{figure}%
\centering
\includegraphics[width=0.45\textwidth]{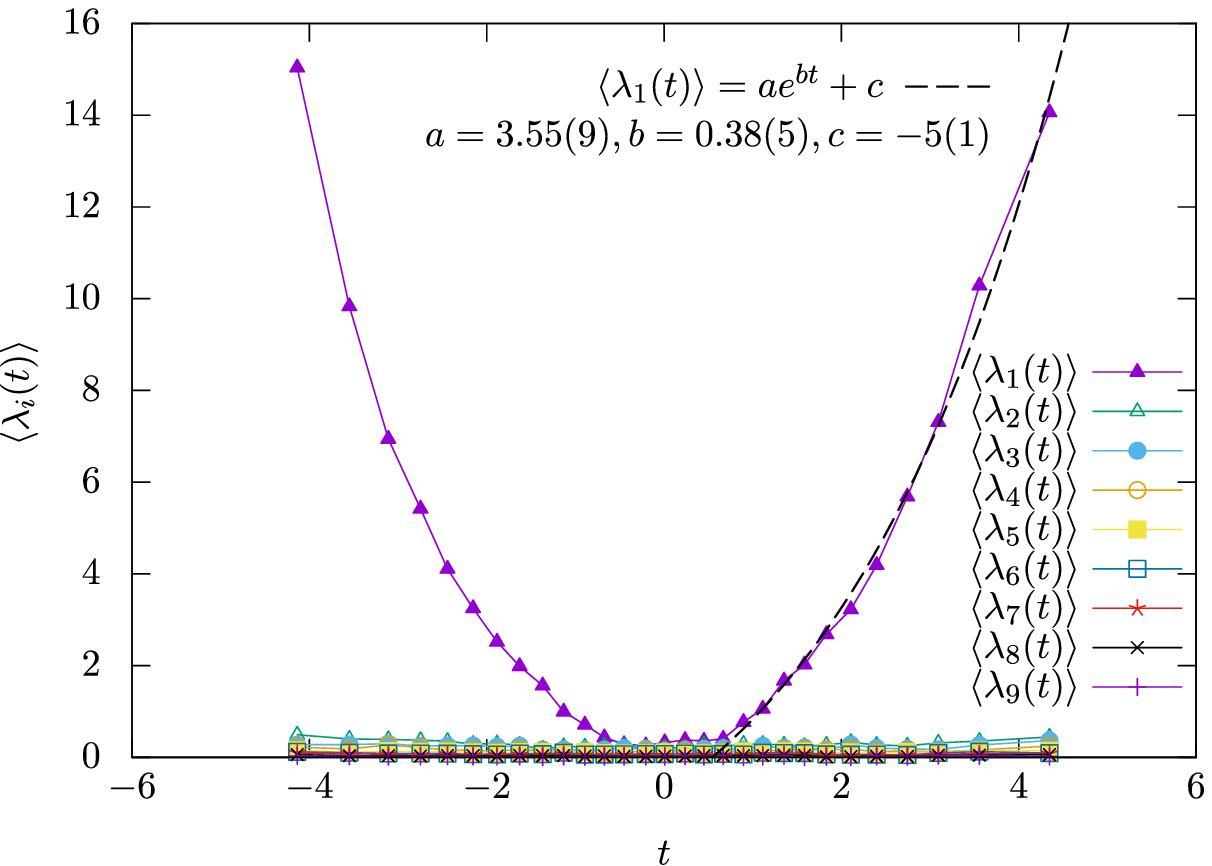}
\includegraphics[width=0.45\textwidth]{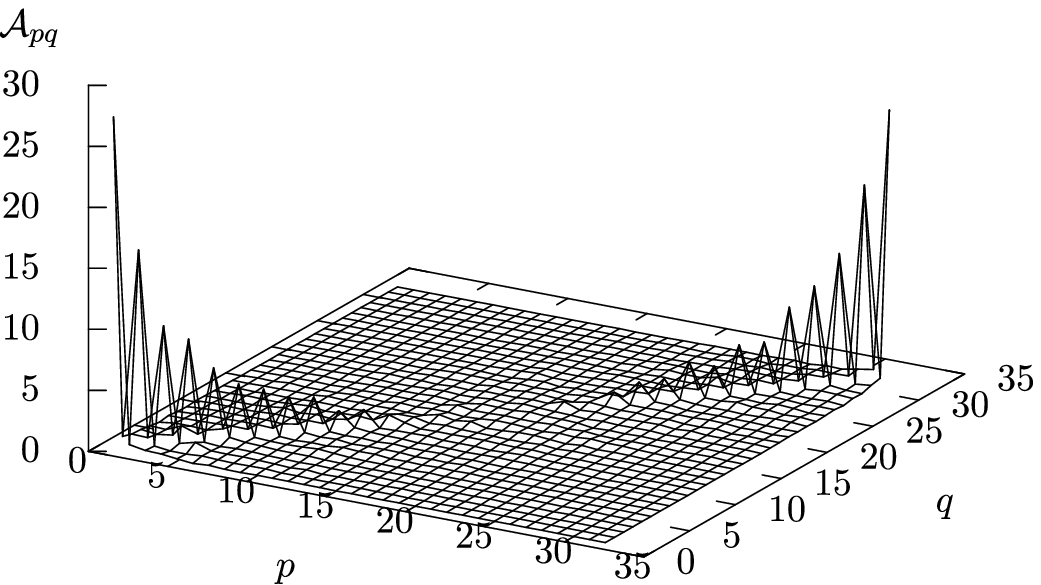}

\caption{(Left) The expectation values $\langle \lambda_I(t) \rangle$ are plotted against the time $t$ for $\gamma=3$. The dashed line represents a fit of $\langle \lambda_1(t) \rangle$ to $\langle \lambda_1(t) \rangle =ae^{bt}+c$ with $a=3.55(9)$, $b=0.38(5)$ and $c=-5(1)$.
(Right) The magnitude $\mathcal{A}_{pq}$ of each element of $A_I$ is plotted against $p$ and $q$ for $\gamma=3$.
Quoted from Fig.~3 of Ref.~\cite{2205_04726}.\label{Tij_blockdiag}}
\end{figure}

In Fig.~\ref{Tij_blockdiag} (Left), we show $\langle \lambda_I(t) \rangle$ for $\gamma=3$, where $\lambda_I(t)$ are the eigenvalues of $T_{IJ}(t)$, 
the order parameter of the SSB of SO(9) symmetry defined in Eq.~(\ref{tensor_tij}). We solve the ninth-order equation 
$\left\langle \det (z \mathbf{1} - T(t)) \right\rangle = \sum_{k=0}^9 \langle c_k \rangle z^k=0$, 
where $\mathbf{1}$ is the $9 \times 9$ unit matrix,
so that the coefficients $c_k$ maintain the holomorphicity with respect to $A_\mu$.
We observe that only one direction expands and the other directions remain small.  
The largest eigenvalue $\langle \lambda_1(t) \rangle$ can be fitted 
nicely to an exponential function. In Fig.~\ref{Tij_blockdiag} (Right) we plot
\begin{align}
\mathcal{A}_{pq} = \frac{1}{9} \sum_{I=1}^9 \vert (A_I)_{pq} \vert^2 \quad (1 \leq p,q \leq N=32)
\end{align}
against $p$ and $q$ for $\gamma=3$. We observe a clear band-diagonal structure with the off-diagonal elements being quite small,
which is important in defining the submatrices \eqref{space_for_t}.
A similar behavior is observed for $\gamma=5,7$, which justifies our choice $n=4$ of the block size for $\gamma \geq 3$ in the present case.
Such band-diagonal structure is not observed for $\gamma \leq 2$.

\section{Summary and outlook}\label{sec13}

In this article, we reviewed the progress in numerical studies on the type IIB matrix model. The numerical simulations of the
Euclidean model using the factorization method and the CLM yield results that are consistent with the SSB of the SO(10) rotational
symmetry to SO(3), in agreement with the GEM. While this is an interesting dynamical property, its relevance to the real world
is unclear.

The Lorentzian version was studied first by using the approximation introduced to avoid the sign problem. This showed the exponential,
followed by the power-law, expansion of three out of nine directions after a critical time. On the other hand, the obtained
three-dimensional space was found to have a singular Pauli-matrix structure, which implies that the emergent space is not smooth.

This motivated the application of the CLM to the Lorentzian model.  Without any cutoffs, the Lorentzian model was found to be equivalent to the
Euclidean model, and that the emergent space-time obtained from the Lorentzian model should be interpreted as Euclidean.

To overcome this situation, we proposed to add a Lorentz-invariant mass term.  Our preliminary results for the
bosonic model are very promising.  When the mass parameter $\gamma$ is large enough, the path integral is dominated by one of the
classical solutions, having Lorentzian signature and expanding behavior. As $\gamma$ becomes smaller, the extent of the emergent time
increases and the emergent space is expanding more at late times.  The expansion at late times is consistent with an exponential
behavior.  The signature of space-time is Lorentzian at late times, while it seems to change to Euclidean
at early times.  We speculate that an expanding space-time with Lorentzian signature emerges at late times in the $\gamma \to +0$
limit after taking the large--$N$ limit.

When space has an expanding behavior, we observe a clear block-diagonal structure, which is important in extracting the
time-evolution from the matrix configurations that we obtain from the model.  We also observe that space appears to be
continuous instead of having the Pauli-matrix structure that was observed previously by using the approximation to avoid the sign
problem.

In the bosonic model, we observed that only one out of nine spatial directions expands.  This may be understood from the
action of the original type IIB matrix model.  Since the spatial directions expand exponentially, the $\text{tr} [A_I,A_J]^2 $ term
becomes dominant.  The fluctuation of this term can be made small by having only one expanding direction.

As a future prospect, it is important to study the impact of the fermionic matrices on the dynamical generation of space-time. We
expect supersymmetry to play an essential role in realizing the expansion of three spatial directions. It is known that $\textrm{Pf
} {\cal M}$ vanishes if we set $A_{\mu}$ to zero except for two of them \cite{9803117,0003223}, which strongly suppresses the
$(1+1)$-dimensional space-time and possibly also $(2+1)$-dimensional space-time considering the exponential expansion of space. As
we have done in the Euclidean model, we have to introduce the fermionic mass term to avoid the singular drift problem in the
CLM. It remains to be seen whether we can reduce the fermionic mass to the extent that enables us to see the effects of supersymmetry
needed to make the space-time $(3+1)$-dimensional.

\section*{Acknowledgements}
T.\;A., K.\;H. and A.\;T. were supported in part by Grant-in-Aid (Nos. 17K05425, 19J10002, and 18K03614, 21K03532, respectively)
from Japan Society for the Promotion of Science.  This research was supported by MEXT as ``Program for Promoting Researches on the
Supercomputer Fugaku'' (Simulation for basic science: from fundamental laws of particles to creation of nuclei, JPMXP1020200105)
and JICFuS.  This work used computational resources of supercomputer Fugaku provided by the RIKEN Center for Computational Science
(Project ID: hp210165, hp220174), and Oakbridge-CX provided by the University of Tokyo (Project IDs: hp200106, hp200130, hp210094,
hp220074) through the HPCI System Research Project.  Numerical computation was also carried out on PC clusters in KEK Computing
Research Center.  This work was also supported by computational time granted by the Greek Research and Technology Network (GRNET)
in the National HPC facility ARIS, under the project IDs SUSYMM and SUSYMM2.  K.\;N.\;A and S.\;K.\;P. were supported in part by a
Program of Basic Research PEVE 2020 (No. 65228700) of the National Technical University of Athens.

\bibliography{sn-IKKT}
\bibliographystyle{sn-mathphys}


\end{document}